\documentclass[aps,prd,twocolumn,superscriptaddress,amssymb,eqsecnum,showpacs,showkeyes,secnumarabic,graphics,floatfix,nofootinbib,tightenlines,longbibliography]{revtex4-1}

\usepackage{graphicx, bm, epsfig, dcolumn, amsmath, amssymb, subcaption, enumerate, academicons}
\usepackage[utf8]{inputenc}
\usepackage[dvipsnames]{xcolor}
\usepackage[breaklinks=true,colorlinks=true,linkcolor=blue,urlcolor=blue,citecolor=MidnightBlue]{hyperref}
\usepackage{booktabs}

\newcommand{\newc}{\newcommand}
\newc{\be}{\begin{equation}}
\newc{\ee}{\end{equation}}
\newc{\ba}{\begin{eqnarray}}
\newc{\ea}{\end{eqnarray}}
\newc{\lcdm}{$\Lambda$CDM }
\newc{\orcid}[1]{\href{https://orcid.org/#1}{\textcolor[HTML]{A6CE39}{\aiOrcid}}}

\begin{document}
\title{Hubble Tension or Distance Ladder Crisis?}
\author{Leandros Perivolaropoulos}\email{leandros@uoi.gr}
\affiliation{Department of Physics, University of Ioannina, GR-45110, Ioannina, Greece}
\email{leandros@uoi.gr}

\date{\today}

\begin{abstract}
We present an up-to-date compilation of published Hubble constant ($H_0$) measurements that are independent of the CMB sound horizon scale. This compilation is split in two distinct groups: A. Distance Ladder Measurements sample comprising of 20 recent measurements, primarily from the past four years, utilizing various rung 2 calibrators and rung 3 cosmic distance indicators. B.One-Step Measurements sample including 33 measurements of $H_0$ that are independent of both the CMB sound horizon scale and the distance ladder approach. These 33 measurements are derived from diverse probes such as Cosmic Chronometers, gamma-ray attenuation, strong lensing, megamasers etc. Statistical analysis reveals a significant distinction between the two samples. The distance ladder-based sample yields a best fit $H_0 = 72.8 \pm 0.5$ km s$^{-1}$ Mpc$^{-1}$ with $\chi^2/dof=0.51$ indicating some correlations. The one-step measurements result in $H_0 = 69.0 \pm 0.48$ km s$^{-1}$ Mpc$^{-1}$ with $\chi^2/dof=1.37$ indicating some internal tension. If two outlier measurements are removed (TDCOSMO.I-2019 known to have systematics and MCP-2020) the best fit of the one step sample reduces to $H_0 = 68.3 \pm 0.5$ km s$^{-1}$ Mpc$^{-1}$ with $\chi^2/dof=0.95$, fully self-consistent and consistent with sound horizon based measurements.  A Kolmogorov-Smirnov test yields a p-value of 0.0001 suggesting that the two samples are fundamentally distinct, with a probability of less than 0.01\% that they are drawn from the same underlying distribution. These findings suggest that the core of the Hubble tension lies not between early and late-time measurements, but between distance ladder measurements and all other $H_0$ determinations. This discrepancy points to either a systematic effect influencing all distance ladder measurements or a fundamental physics anomaly affecting at least one rung of the distance ladder. 
\end{abstract}

\maketitle

\section{Introduction}

The Hubble constant (\(H_0\)) serves as a fundamental metric for quantifying the rate at which the Universe is expanding. Accurate determination of \(H_0\) is essential for elucidating the Universe's age, composition, and ultimate fate. Despite remarkable progress in observational astrophysics and cosmology, a notable and persistent discrepancy, commonly referred to as the "Hubble tension"\cite{Perivolaropoulos:2021jda,Abdalla:2022yfr,DiValentino:2021izs,Bernal:2016gxb,Hu:2023jqc,Efstathiou:2024dvn,Efstathiou:2021ocp,Khalife:2023qbu,Tully:2023bmr,Dainotti:2023yrk,Abdalla:2022yfr,Shah:2021onj,Schoneberg:2021qvd,Bousis:2024rnb},  remains between the \(H_0\) values deduced between different cosmological probes. It is common perception that measurements of $H_0$ that use the sound horizon at recombination as a standard ruler distance calibrator (early time measurements) lead to a different (lower) value of \(H_0 = 67.4 \pm 0.5\) km/s/Mpc \cite{Planck:2018vyg} than most other measurements which favor a value consistent with the latest SH0ES measurement favoring an elevated \(H_0\) value of \(73.04 \pm 1.04\)km/s/Mpc\cite{Riess:2021jrx}). It is therefore believed by a large sector of the scientific community that an adjustment of the prerecombination physics in such a way that the recombination sound horizon becomes smaller, could lead to a larger measured value of $H_0$ and thus to consistency between the early time sound horizon based measurements and all other measurements of $H_0$\cite{Poulin:2018cxd,Kamionkowski:2022pkx,Simon:2022adh,Braglia:2020bym,Niedermann:2020dwg,Smith:2020rxx,Rezazadeh:2022lsf,Braglia:2020auw,Brax:2013fda,Abadi:2020hbr,Clifton:2011jh,Lin:2018nxe,DiValentino:2015bja,CANTATA:2021ktz,Rossi:2019lgt,Braglia:2020iik,FrancoAbellan:2023gec,Seto:2021xua,Sakstein:2019fmf,Vagnozzi:2019ezj,Smith:2020rxx}. This approach if faced by a wide range of challenges\cite{Jedamzik:2020zmd,Jedamzik:2020zmd,Vagnozzi:2023nrq,Vagnozzi:2021gjh,Hill:2020osr,Goldstein:2023gnw,Fondi:2022xsf}. It is based on the following assumption: {\it Most measurements of $H_0$ that are not based on the Cosmic Microwave Backgroud (CMB) sound horizon as a standard ruler favor a higher value of $H_0$ consistent with the SH0ES measurement which is the most precise local measurement of $H_0$.} 

This assumption is perceived to be correct because the most precise local measurements of $H_0$ are indeed fully consistent with the SH0ES measurement. These measurements which are highly precise but not necessarily accurate, are based on {\it distance ladder methods}. 

A crucial assumption of distance ladder methods is that both the astrophysical environment and the physical laws\cite{Marra:2021fvf,Alestas:2020zol,Perivolaropoulos:2021bds} are consistent  among the three rungs used in the distance ladder implementation. Thus, in the context of testing this assumpotion, the following questions arise:
\begin{itemize}
    \item What are the observational probes that can lead to measurement of $H_0$ through a one step process that are independent of both the sound horizon scale and of the distance ladder rung approach?
    \item Is there a trend for the values of $H_0$ favored by these probes and if yes, is the favored range of $H_0$ more consistent with sound horizon based measurements or with distance ladder based measurements of $H_0$?
    \item Are the distance ladder based measurements of $H_0$ consistent with each other and with the latest SH0ES measurement? 
\end{itemize}

The main goal of the present analysis is to address these questions using an extensive up-to-date compilation of recent measurements of $H_0$ (most published during the past 5 years with emphasis in the last 3 years) which are independent of the sound horizon scale and include both distance ladder measurements and one step measurements that involve no rungs. We thus split the measurements of $H_0$ that are independent of the sound horizon scale in two groups: 1. Distance ladder rung based measurements 2. One step measurements that involve direct distance or $H(z)$ probes and are independent of the recombination sound horizon scale. We then test the statistical properties of the $H_0$ measurements in each group in order to address the above questions. This is the first time that such a statistical meta-analysis is attempted since previous corresponding studies have not separated the two groups of measurements and have also included many early outdated measurements\cite{Wang:2023bxf,Lopez-Corredoira:2022qdo,Cao:2023eja,Verde:2023lmm}.

The structure of this paper is the following: In the next section we review the methods for measuring $H_0$ and classify them into sound horizon based methods,  distance ladder methods that require calibrators and rung structure and one step methods that are free from local astrophysics and local calibrators. In section III we present our extensive compilation of the currently available $H_0$ measurements in each one of the later two groups  (distance ladder and one step measurements independent of sound horizon). We also present the statistical properties of the two groups and their consistency with each other and with the $H_0$ measurements based on the CMB sound horizon scale standard ruler. Finally in section IV we summarize our main results and their implications. We also discuss possible future extensions of the current analysis. 

\section{Methods for the measurement of $H_0$}
Methods for the measurement of $H_0$ may be classified in three broad classes: Methods that use the CMB sound horizon standard ruler as a distance calibrator, distance ladder methods and one step methods independent of the sound horizon scale input. In this section we briefly review the main representative methods of each of of these three classes which are used in the present analysis.

\subsection{CMB Sound Horizon Standard Ruler Methods}
The CMB provides a powerful standard ruler in cosmology—the sound horizon at recombination. This standard ruler is pivotal for inferring the Hubble constant, \(H_0\), through observations of the CMB power spectrum's acoustic peaks and the imprint of baryon acoustic oscillations (BAO) in the distribution of galaxies.

\subsubsection{Sound Horizon at Recombination}

The sound horizon at recombination, \(r_s\), represents the maximum distance that acoustic waves could have traveled in the primordial photon-baryon fluid by the time of recombination\cite{Wald1984,Weinberg1972}. This scale is imprinted as the characteristic angular scale of fluctuations observed in the CMB power spectrum. Mathematically, it is expressed as:

\begin{equation}
    r_s = \int_{z_{\text{rec}}}^{\infty} \frac{c_s(z)}{H(z)} \, dz,
    \label{rshz}
\end{equation}

where \(c_s(z)\) is the sound speed in the photon-baryon fluid, \(H(z)\) is the Hubble parameter as a function of redshift, and \(z_{\text{rec}}\) is the redshift at recombination. The sound speed \(c_s\) is lower than the speed of light due to the inertia of baryons, which are coupled to photons by Thomson scattering until recombination.

In the standard \(\Lambda\)CDM model, the sound speed can be approximated as:

\begin{equation}
    c_s(z) = \frac{c}{\sqrt{3(1 + R(z))}},
\end{equation}
where \(R(z)\) is the ratio of the baryon to photon momentum density\cite{Eisenstein:1997ik}:

\begin{equation}
    R(z) = \frac{3\rho_b}{4\rho_\gamma} = \frac{3\Omega_b}{4\Omega_\gamma}(1+z)^{-1}.
\end{equation}

Using the latest Planck 2018\cite{Planck:2018vyg} results for the \(\Lambda\)CDM model, which give \(\Omega_b h^2 = 0.02237\) and \(\Omega_m h^2 = 0.1430\), and assuming \(z_{\text{rec}} \approx 1090\), the calculated value of \(r_s\) is approximately:

\begin{equation}
    r_s \approx 147.09 \pm 0.26 \text{ Mpc}.
\end{equation}

This value of \(r_s\) is crucial in determining \(H_0\) from CMB and BAO measurements.

The sound horizon can be modified by changing the physics before recombination. Some proposed modifications include:

1. Early Dark Energy (EDE)\cite{Poulin:2023lkg,Poulin:2018cxd,Kamionkowski:2022pkx,Simon:2022adh,Braglia:2020bym,Niedermann:2020dwg,Smith:2020rxx,Rezazadeh:2022lsf,Agrawal:2019lmo}: Introducing a new component of dark energy that is significant before recombination can increase the expansion rate, reducing the time available for sound waves to propagate and thus decreasing \(r_s\).

2. Modified Neutrino Physics\cite{Seto:2021xua,Sakstein:2019fmf,Vagnozzi:2019ezj}: Changing the number of effective neutrino species or their interactions can alter the expansion history and the sound speed, affecting \(r_s\).

3. Varying Fundamental Constants\cite{Hart:2021kad}: Allowing fundamental constants like the fine-structure constant to vary over time can change the recombination history and thus \(r_s\).

4. Non-standard Recombination\cite{Lynch:2024gmp,Lee:2022gzh}: Modifications to the recombination process itself, such as energy injection from dark matter annihilation, can change \(z_{\text{rec}}\) and consequently \(r_s\).

5. Modified Gravity\cite{Braglia:2020auw,Brax:2013fda,Abadi:2020hbr,Clifton:2011jh,Lin:2018nxe,DiValentino:2015bja,CANTATA:2021ktz,Rossi:2019lgt,Braglia:2020iik,FrancoAbellan:2023gec,Pogosian:2021mcs}: Alterations to general relativity in the early universe can change the expansion history and thus \(r_s\).

These modifications aim to reduce \(r_s\), which would allow for a higher \(H_0\) value from CMB and BAO measurements, potentially alleviating the Hubble tension. However, each of these proposals faces challenges in maintaining consistency with other cosmological observations and often requires fine-tuning\cite{Jedamzik:2020zmd,Jedamzik:2020zmd,Vagnozzi:2023nrq,Vagnozzi:2021gjh,Hill:2020osr,Goldstein:2023gnw,Fondi:2022xsf}.

\subsubsection{Measuring \(H_0\) from the CMB}
The angular scale of the first acoustic peak in the CMB power spectrum, \(\theta_s\), is inversely proportional to the sound horizon at the surface of last scattering\cite{Efstathiou:2000xt,WMAP:2003elm,Planck:2018vyg,ACT:2020frw}:
\begin{equation}
    \theta_s = \frac{r_s}{D_A(z_{\text{rec}})},
    \label{thetas}
\end{equation}
where \(D_A(z_{\text{rec}})\) is the angular diameter distance to the surface of last scattering. In the context of the standard \lcdm model, the angular diameter distance is related to the Hubble constant \(H_0\) through the following equation:

\begin{equation}
    D_A(z_{\text{rec}}) = \frac{c}{H_0(1+z_{\text{rec}})}\int_0^{z_{\text{rec}}} \frac{dz'}{\sqrt{\Omega_m(1+z')^3 + \Omega_\Lambda}},
\end{equation}
where \(c\) is the speed of light, \(z_{\text{rec}}\) is the redshift of recombination, \(\Omega_m\) is the matter density parameter, and \(\Omega_\Lambda\) is the cosmological constant density parameter.

Precise measurements of \(\theta_s\) from CMB experiments like Planck, combined with this model for \(D_A\) that depends on \(H_0\) and other cosmological parameters, allow for the inference of the Hubble constant. However, it's important to note that \(H_0\) is degenerate with \(r_s\) in this measurement.

This degeneracy arises because the CMB primarily constrains the combination \(r_s H_0\) rather than either parameter individually. To see this, we can rewrite the equation \eqref{thetas} for \(\theta_s\) as:

\begin{equation}
    \theta_s = \frac{r_s H_0}{c(1+z_{\text{rec}})}\left(\int_0^{z_{\text{rec}}} \frac{dz'}{\sqrt{\Omega_m(1+z')^3 + \Omega_\Lambda}}\right)^{-1}
\end{equation}

In this form, it's clear that a larger \(r_s\) could be compensated by a smaller \(H_0\), or vice versa, while still producing the same observed \(\theta_s\). This degeneracy means that additional cosmological probes or assumptions about the early universe are needed to break the degeneracy and determine \(H_0\) uniquely from CMB data. In the context of \lcdm $r_s$ is calculated as described in the previous subsection. However, in the context of modified cosmological models $r_s$ will vary as the expansion history before recombination $H(z)$ changes (see equation \eqref{rshz}).

\subsubsection{Connecting BAO and the CMB}

Building on the CMB measurements discussed in the previous subsection, Baryon Acoustic Oscillations (BAO) provide an independent confirmation of the sound horizon scale\cite{SDSS:2005xqv,BOSS:2014hhw,eBOSS:2020lta,Favale:2024sdq}. The BAO feature serves as a standard ruler in the late-time universe, manifesting as a peak in the correlation function of galaxies. This peak corresponds to the same comoving scale $r_s$ imprinted in the CMB.

The BAO measurements typically constrain the combination of the sound horizon and a distance scale. For example, in terms of the volume-averaged distance $D_V(z)$, we have:

\begin{equation}
    \frac{D_V(z)}{r_s} = \frac{1}{r_s}\left[(1+z)^2 D_A^2(z) \frac{cz}{H(z)}\right]^{1/3},
\end{equation}
where z is the redshift of the galaxy survey. The left-hand side of this equation is what BAO surveys directly constrain.

Similar to the CMB case, we can see that $r_s$ and $H_0$ are degenerate in BAO measurements. To illustrate this, let's consider how $H_0$ enters into $D_V(z)$:

\begin{equation}
    D_V(z) = \left[(1+z)^2 \left(\frac{c}{H_0}\int_0^z \frac{dz'}{E(z')}\right)^2 \frac{cz}{H_0E(z)}\right]^{1/3},
\end{equation}
where $E(z) = H(z)/H_0$ is the normalized Hubble parameter.

Now, we can rewrite the BAO constraint as:

\begin{equation}
    \frac{D_V(z)}{r_s} = \frac{c}{r_s H_0} \left[(1+z)^2 \left(\int_0^z \frac{dz'}{E(z')}\right)^2 \frac{z}{E(z)}\right]^{1/3}
\end{equation}

In this form, we can see that BAO measurements primarily constrain the combination $r_s H_0$, similar to what we saw with CMB measurements. A larger $r_s$ could be compensated by a smaller $H_0$, or vice versa, while still satisfying the BAO constraint.

This degeneracy between $r_s$ and $H_0$ in both CMB and BAO measurements highlights why these early-universe probes, while precise, cannot uniquely determine $H_0$ without additional assumptions or complementary data. It also underscores the importance of late-time, direct measurements of $H_0$ that don't rely on the sound horizon scale, as these can potentially break the degeneracy and help resolve the Hubble tension.

These methods, grounded in the physics of the early universe, provide robust constraints on cosmological parameters but are sensitive to assumptions about the cosmological model, primarily the composition and evolution of the universe's energy density. As such, they are crucial in exploring and potentially resolving tensions in the measured values of $H_0$ observations.

In addition to $D_V(z)$, BAO measurements can also constrain the comoving distance $D_M(z)$. The comoving distance is related to the angular diameter distance by $D_M(z) = (1+z)D_A(z)$ . In terms of $H_0$, $D_M$ can be expressed as:

\begin{equation}
    D_M(z) = \frac{c}{H_0} \int_0^z \frac{dz'}{E(z')}
\end{equation}
where $E(z) = H(z)/H_0$ as before. BAO surveys often measure the combination $r_s/D_M(z)$, which again demonstrates the degeneracy between $r_s$ and $H_0$:

\begin{equation}
    \frac{r_s}{D_M(z)} = \frac{r_s H_0}{c} \left(\int_0^z \frac{dz'}{E(z')}\right)^{-1}
\end{equation}

This formulation clearly shows that BAO measurements of $D_M$, like those of $D_V$, primarily constrain the product $r_s H_0$ rather than either parameter individually.

It's important to note that while we've been discussing the sound horizon at recombination ($r_s$), BAO measurements are actually sensitive to a slightly different scale: the sound horizon at the drag epoch ($r_d$). The drag epoch occurs when baryons decouple from photons, which happens slightly later than recombination. The drag scale $r_d$ is defined as the comoving distance a sound wave can travel from the beginning of the universe until the drag epoch:

\begin{equation}
r_d = \int_0^{t_d} \frac{c_s(t)}{\sqrt{3}(1+z(t))} dt,
\end{equation}
where $t_d$ is the time of the drag epoch and $c_s(t)$ is the sound speed in the photon-baryon fluid. The drag scale $r_d$ is typically about 2\% larger than $r_s$. This distinction is crucial because it's the baryons, not the photons, that seed the matter distribution we observe in galaxy surveys. Therefore, the BAO feature in the galaxy distribution corresponds to $r_d$ rather than $r_s$. However, the two scales are very closely related, and the degeneracy with $H_0$ applies equally to both. In precise analyses, it's important to use $r_d$ for BAO calculations, while $r_s$ remains the relevant scale for CMB analyses.

\subsection{Distance Ladder Approaches}
The distance ladder approach\cite{HST:2000azd,Riess:2016jrr,Freedman:2024eph,Riess:2021jrx} in cosmology is a methodical technique used to determine cosmological distances across the universe, spanning from nearby stars to distant galaxies. This method is structured into three sequential "rungs," each building upon the accuracy of the previous to extend further into the cosmos.

\subsubsection{First Rung: Calibration of Local Distance Indicators}

The primary objective of the first rung in the cosmic distance ladder is to calibrate nearby distance indicators, specifically Cepheid variables\cite{Feast:1997zz,Madore:1991yf}, Tip of the Red Giant Branch (TRGB) stars\cite{Lee:1993jb,Rizzi:2007ni}, Mira variables\cite{Glass:1981,Feast:1989} and J-region Asymptotic Giant Branch (JAGB) stars\cite{Madore:2020yqv}. This calibration process involves determining the precise calibration parameters for each indicator using direct geometric distance measurements.

\paragraph{Cepheid Variables and Their Calibration}

Cepheid variables are a class of pulsating stars that play a crucial role in the cosmic distance ladder. Named after the prototype star $\delta$ Cephei, these stars are yellow supergiants that undergo regular variations in their brightness. The key to their importance in astronomy lies in the tight relationship between their pulsation period and luminosity, first discovered by Henrietta Swan Leavitt in 1908 \cite{Leavitt1908}.

Cepheids pulsate due to a mechanism known as the kappa mechanism, where ionization zones in the star's atmosphere drive periodic expansions and contractions. This pulsation leads to regular changes in the star's brightness and temperature, with periods typically ranging from a few days to months.

The Period-Luminosity (PL) relation, also known as the Leavitt Law, states that the intrinsic luminosity of a Cepheid is directly related to its pulsation period. This relationship allows astronomers to use Cepheids as ``standard candles'' for measuring cosmic distances. However, modern calibrations have refined this relationship to include metallicity effects, leading to the Period-Luminosity-Metallicity (PLZ) relation:

\begin{equation}
    M_\lambda = a_\lambda \log P + b_\lambda + c_\lambda [\text{Fe/H}] + d_\lambda
    \label{eq:cepheidplz}
\end{equation}

where:
\begin{itemize}
    \item $M_\lambda$ is the absolute magnitude in wavelength band $\lambda$
    \item $P$ is the pulsation period in days
    \item The symbol [Fe/H] represents the metallicity of the star. Fe/H represents the logarithm of the ratio of iron to hydrogen abundance in a star, compared to the same ratio in the Sun. It's expressed mathematically as:
\begin{equation}
[\text{Fe/H}] = \log_{10}\left(\frac{N_\text{Fe}}{N_\text{H}}\right)\text{star} - \log_{10}\left(\frac{N_\text{Fe}}{N_\text{H}}\right)_\text{Sun}
\end{equation}
Where $N_\text{Fe}$ and $N_\text{H}$ are the number densities of iron and hydrogen atoms respectively.
    \item $a_\lambda$, $b_\lambda$, $c_\lambda$, and $d_\lambda$ are calibration parameters
\end{itemize}

The calibration of this relation is crucial for accurate distance measurements. Modern calibrations (e.g., \cite{Riess:2021jrx,Perivolaropoulos:2022khd}) utilize a multi-pronged approach:

\begin{enumerate}
    \item \textbf{Parallax Measurements:} The European Space Agency's Gaia mission has provided precise parallax measurements for Galactic Cepheids \cite{Gaia:2018ydn}, allowing direct distance determinations for nearby Cepheids.
    
    \item \textbf{Anchor Galaxies:} Geometric distances to nearby galaxies hosting Cepheids provide additional calibration points. Key examples include:
    \begin{itemize}
        \item NGC 4258, with a precise maser-based distance
        \item The Large Magellanic Cloud (LMC), with distances determined from eclipsing binary systems
    \end{itemize}
    
    \item \textbf{Multi-wavelength Observations:} Cepheids are observed across multiple wavelength bands to constrain extinction and metallicity effects. Common bands include:
    \begin{itemize}
        \item Optical: B (445 nm), V (551 nm), R (658 nm), I (806 nm)
        \item Near-infrared: J (1.25 $\mu$m), H (1.65 $\mu$m), K (2.17 $\mu$m)
        \item Space-based: HST WFC3 F160W (1.60 $\mu$m)
    \end{itemize}
    Near-infrared bands are particularly valuable as they are less affected by extinction and metallicity variations.
    
    \item \textbf{Period Range:} Calibrations typically focus on fundamental mode Cepheids with periods between 3 and 100 days to ensure consistency and reliability.
\end{enumerate}

The calibration process involves simultaneously fitting data from these various sources, accounting for systematic effects such as crowding, metallicity gradients, and extinction. Advanced statistical techniques, including Bayesian hierarchical models, are often employed to handle the complex interdependencies in the data \cite{Riess:2021jrx}.

By refining the PLZ relation through these meticulous calibrations, astronomers can use Cepheid variables to measure distances to galaxies up to about 40 Mpc, providing a crucial rung in the cosmic distance ladder and playing a vital role in the determination of the Hubble constant.

\paragraph{The Tip of the Red Giant Branch (TRGB) Method and Its Calibration}

The Tip of the Red Giant Branch (TRGB) method is a powerful technique for measuring extragalactic distances, complementing the Cepheid variable approach in the cosmic distance ladder. This method utilizes the predictable maximum luminosity of red giant stars in their final evolutionary stages.

Red giant stars are evolved low- to intermediate-mass stars that have exhausted the hydrogen in their cores. As these stars ascend the red giant branch, they reach a maximum luminosity just before the onset of helium fusion in their cores. This maximum luminosity, corresponding to the brightest red giant stars, creates a sharp cut-off in the luminosity function of the red giant population, which is easily identifiable in color-magnitude diagrams \cite{DaCosta:1990}.

The TRGB method was first proposed as a distance indicator by Baade in the 1940s \cite{Baade:1944}, but it wasn't until the 1990s that it was developed into a precise tool for extragalactic distance measurements \cite{Lee:1993}. The key advantage of the TRGB method is that it relies on old, low-mass stars present in all galaxy types, unlike Cepheid variables which are only found in star-forming regions.

The calibration of the TRGB method focuses on determining the absolute magnitude of the TRGB, primarily in the I-band ($\sim$800 nm). The I-band is preferred because the TRGB magnitude is nearly constant for old, metal-poor populations in this wavelength range. The calibration can be expressed as:

\begin{equation}
    M_I^{\text{TRGB}} = a [\text{Fe/H}] + b
    \label{eq:trgbcal}
\end{equation}
where $M_I^{\text{TRGB}}$ is the absolute magnitude of the TRGB in the I-band, [Fe/H] is the metallicity, and $a$ and $b$ are calibration parameters \cite{Freedman:2020dne, Li:2024gib}.

The calibration process involves several key components:

\begin{enumerate}
    \item \textbf{Photometric Systems:} While the I-band is most commonly used, observations in multiple bands (e.g., V, R, J, H, K) help constrain extinction and metallicity effects. The Hubble Space Telescope's ACS F814W filter is often used for space-based observations.
    
    \item \textbf{Metallicity Corrections:} The dependence on metallicity, represented by the $a[\text{Fe/H}]$ term, is crucial for accurate calibration, especially for metal-rich populations \cite{Jang:2021}.
    
    \item \textbf{Anchor Galaxies:} Similar to Cepheid calibration, the TRGB method uses anchor galaxies with independently measured distances. Common anchors include:
    \begin{itemize}
        \item The Large Magellanic Cloud (LMC)
        \item NGC 4258 (with maser-based distance)
        \item Milky Way globular clusters with precise parallax measurements
    \end{itemize}
    
    \item \textbf{Detection Algorithms:} Sophisticated edge-detection algorithms are employed to precisely locate the TRGB in color-magnitude diagrams \cite{Makarov:2006}.
    
    \item \textbf{Population Effects:} Careful consideration of the stellar population characteristics, including age and metallicity distribution, is necessary for accurate calibration \cite{Serenelli:2017}.
\end{enumerate}

Recent calibrations of the TRGB method have achieved precisions comparable to or even exceeding those of Cepheid-based measurements. For instance, \citet{Freedman:2020dne} reported a calibration of $M_I^{\text{TRGB}} = -4.05 \pm 0.02$ (stat) $\pm 0.039$ (sys) mag, while \citet{Li:2024gib} provided an updated calibration incorporating the latest Gaia data.

The TRGB method can be applied to distances up to about 20 Mpc, making it a valuable tool for calibrating secondary distance indicators like Type Ia supernovae. Its independence from Cepheid-based measurements provides an important cross-check in the cosmic distance ladder and in determinations of the Hubble constant.

\paragraph{Mira Variables and Their Calibration}

Mira variables, named after the prototype star Mira (o Ceti), are a class of long-period variable stars that play an increasingly important role in the cosmic distance ladder. These stars are cool, highly evolved stars on the asymptotic giant branch (AGB) of the Hertzsprung-Russell diagram, characteristically old ($>1$ Gyr) and of low to intermediate mass (0.8-8 $M_\odot$) \cite{Whitelock:2012}.

Mira variables undergo large-amplitude pulsations with periods typically ranging from 100 to 1000 days, causing their brightness to vary by several magnitudes. These pulsations are believed to be driven by a combination of changes in opacity and convection in the stellar envelope \cite{Wood:2015}.

The potential of Mira variables as distance indicators was recognized in the early 20th century, with initial period-luminosity relationships established by Gerasimovič (1928) and Spine (1948) \cite{Gerasimovic:1928,Spine:1948}. However, it wasn't until the advent of near-infrared astronomy in the 1980s that their full potential as standard candles began to be realized \cite{Glass:1981}.

The calibration of Mira variables for distance measurement primarily relies on their Period-Luminosity (PL) relation, which is most commonly expressed in the near-infrared K-band ($\sim$2.2 $\mu$m):

\begin{equation}
    M_K = a \log P + b
    \label{eq:miracal}
\end{equation}
where $M_K$ is the absolute magnitude in the K-band, $P$ is the pulsation period in days, and $a$ and $b$ are calibration parameters \cite{Whitelock:2008mz,Huang:2024exg}.

The calibration process for Mira variables involves several key aspects:

\begin{enumerate}
    \item \textbf{Wavelength Selection:} While the K-band is most commonly used due to its reduced sensitivity to metallicity and circumstellar dust, observations in multiple infrared bands (e.g., J at 1.25 $\mu$m, H at 1.65 $\mu$m, L' at 3.8 $\mu$m) are often employed to constrain extinction and chemical composition effects \cite{Yuan:2017}.
    
    \item \textbf{Period Determination:} Accurate period determination requires long-term monitoring, typically over several years, to account for cycle-to-cycle variations and potential period changes \cite{Templeton:2005}.
    
    \item \textbf{Chemical Composition:} Mira variables are categorized into oxygen-rich (O-rich) and carbon-rich (C-rich) types based on their surface chemistry. These types follow slightly different PL relations, necessitating careful classification \cite{Ita:2004}.
    
    \item \textbf{Anchor Galaxies:} Calibration of the PL relation relies on Mira variables in well-studied environments with independent distance measurements. Key calibrators include:
    \begin{itemize}
        \item The Large and Small Magellanic Clouds
        \item Galactic Miras with parallax measurements from Hipparcos and Gaia
        \item Miras in Galactic globular clusters
    \end{itemize}
    
    \item \textbf{Metallicity Effects:} While less pronounced than for Cepheids, metallicity can affect the PL relation. Some calibrations include a metallicity term, especially for applications in diverse galactic environments \cite{Whitelock:2008mz}.
    
    \item \textbf{Pulsation Mode:} Miras are believed to pulsate primarily in the fundamental mode, but overtone pulsators exist. Proper mode identification is crucial for accurate calibration \cite{Wood:2015}.
\end{enumerate}

Recent calibrations have demonstrated the power of Mira variables as distance indicators. For instance, \citet{Huang:2024exg} provided an updated calibration incorporating the latest Gaia data, achieving precisions competitive with other primary distance indicators. \citet{Whitelock:2008mz} reported a K-band PL relation with a slope of $a \approx -3.51$ and an intrinsic scatter of $\sim$0.13 mag for O-rich Miras in the LMC.

Mira variables offer several advantages as distance indicators:
\begin{itemize}
    \item They are brighter than Cepheids in the infrared, potentially extending the reach of primary distance measurements.
    \item They are present in all types of galaxies, including early-type galaxies lacking young Cepheid populations.
    \item Their long periods make them less susceptible to aliasing in sparsely sampled datasets.
\end{itemize}

As observational techniques and calibrations continue to improve, Mira variables are becoming an increasingly valuable tool in the cosmic distance ladder, complementing and cross-checking other methods in the ongoing effort to refine measurements of the Hubble constant and other cosmological parameters.

\paragraph{J-region Asymptotic Giant Branch (JAGB) Stars and Their Calibration as Distance Indicators}

J-region Asymptotic Giant Branch (JAGB) stars have emerged as a promising new standard candle for the cosmic distance ladder, offering a complementary approach to established methods such as Cepheid variables and the Tip of the Red Giant Branch (TRGB) \cite{Madore:2020yqv,Lee:2024qzr,Freedman:2024eph}. These stars represent a specific subset of carbon-rich Asymptotic Giant Branch (AGB) stars, characterized by their location in color-magnitude diagrams and their near-constant luminosity in the near-infrared J-band.

JAGB stars are evolved, low- to intermediate-mass stars (typically 1.5-4 $M_\odot$) in the thermally pulsing AGB phase. During this stage, these stars undergo periodic helium shell flashes, leading to the dredge-up of carbon-rich material to the stellar surface. This process results in a transformation from oxygen-rich to carbon-rich composition when the C/O ratio exceeds unity \cite{Karakas:2014}.

The key property that makes JAGB stars valuable as distance indicators is the observed constancy of their luminosity in the J-band (centered at $\sim$1.25 $\mu$m). This phenomenon is attributed to a combination of factors:

\begin{itemize}
    \item The core mass-luminosity relation for AGB stars
    \item Temperature regulation due to molecular opacity effects in carbon-rich atmospheres
    \item The specific characteristics of the J-band, which is less affected by molecular absorption than other near-infrared bands
\end{itemize}

The potential of carbon stars as distance indicators was recognized as early as the 1980s \cite{Richer:1981}, but it was the work of Weinberg and Nikolaev in 2001 that first identified the distinct "J" region in color-magnitude diagrams of the Large Magellanic Cloud (LMC) \cite{Weinberg:2001}. However, it wasn't until the comprehensive study by \citet{Madore:2020yqv} that JAGB stars were formally proposed and developed as a new standard candle for extragalactic distances.

The calibration of JAGB stars as distance indicators is based on the empirical finding that their absolute magnitude in the J-band is approximately constant:

\begin{equation}
M_J = \text{constant}
\label{eq:jagbcal}
\end{equation}

where $M_J$ is the absolute magnitude in the J-band.

The calibration process involves several key steps:

\begin{enumerate}
    \item \textbf{Photometric System:} While the J-band (1.25 $\mu$m) is the primary calibration band, observations in multiple near-infrared bands (e.g., J, H at 1.65 $\mu$m, and K$_s$ at 2.15 $\mu$m) are typically used to isolate the JAGB population through color selection.
    
    \item \textbf{Color Selection:} JAGB stars are identified using color cuts, typically in the (J-K$_s$) vs. K$_s$ color-magnitude diagram. The exact color range may vary slightly between studies but generally falls within $1.4 \lesssim (J-K_s) \lesssim 2.0$ \cite{Madore:2020yqv}.
    
    \item \textbf{Luminosity Function:} The JAGB luminosity function in the J-band is analyzed to determine the characteristic magnitude, often using techniques such as Gaussian fitting or edge detection algorithms.
    
    \item \textbf{Anchor Galaxies:} Similar to other distance indicators, JAGB calibration relies on observations in galaxies with well-established distances. Primary calibrators include:
    \begin{itemize}
        \item The Large and Small Magellanic Clouds
        \item M31 (Andromeda) and M33
        \item Galaxies with maser distances (e.g., NGC 4258)
    \end{itemize}
    
    \item \textbf{Metallicity Effects:} While the J-band magnitude of JAGB stars appears to be less sensitive to metallicity than other indicators, potential metallicity effects are still an active area of research \cite{Lee:2024qzr}.
    
    \item \textbf{Population Effects:} The presence and characteristics of JAGB stars depend on the age and star formation history of the stellar population, necessitating careful consideration in diverse galactic environments.
\end{enumerate}

Recent calibrations have demonstrated the promise of the JAGB method. For instance, \citet{Madore:2020yqv,Freedman:2024eph} reported an absolute J-band magnitude of $M_J = -6.22 \pm 0.04$ mag for JAGB stars based on observations in the LMC, with a typical dispersion of $\sim$0.23 mag for individual stars.

The JAGB method offers several advantages as a distance indicator:

\begin{itemize}
    \item \textbf{Extended reach:} JAGB stars are typically $\sim$1 magnitude brighter than the TRGB in the near-infrared, potentially extending the range of primary distance measurements.
    \item \textbf{Single-epoch observations:} Unlike variable stars, JAGB stars can be detected and measured with a single near-infrared observation.
    \item \textbf{Ease of identification:} They are readily identifiable by their distinctive colors and magnitudes in near-infrared color-magnitude diagrams.
    \item \textbf{Reduced extinction:} Near-infrared observations are less affected by dust extinction compared to optical wavelengths.
    \item \textbf{Abundance in various galaxy types:} JAGB stars are present in all galaxies with intermediate-age stellar populations, including early-type galaxies where Cepheids are absent.
    \item \textbf{Independent cross-check:} Their use provides an additional, independent method to verify distances obtained through other techniques.
\end{itemize}

The JAGB method has been successfully applied to measure distances to nearby galaxies and galaxy groups, including the M81 group \cite{Madore:2020yqv} and the Leo I group \cite{Lee:2024qzr}. As observational techniques and calibrations continue to improve, JAGB stars are poised to play an increasingly important role in the cosmic distance ladder, contributing to the ongoing refinement of the Hubble constant and other cosmological parameters.

\subsubsection{Second Rung: Calibration of Type Ia Supernovae}

The primary objective of the second rung is to calibrate Type Ia Supernovae (SnIa) using the local distance indicators calibrated in the first rung. This calibration is crucial for extending distance measurements to cosmological scales.
The calibration of SnIa involves determining their standardized absolute magnitude, which is typically expressed in the B-band, $M_B^{\text{SN}}$. The process includes:

1. Identifying host galaxies containing both SnIa and calibrated distance indicators (primarily Cepheids or TRGB).

2. Measuring distances to these galaxies using the calibrated indicators from the first rung.

3. Standardizing SN Ia luminosities using the Tripp relation \cite{Tripp:1997wt}:

   \begin{equation}
       m_B^{\text{corr}} = m_B + \alpha x_1 - \beta c
       \label{eq:tripp}
   \end{equation}
where $m_B^{\text{corr}}$ is the corrected B-band magnitude, $m_B$ is the observed peak magnitude, $x_1$ is a stretch parameter, $c$ is a color parameter, and $\alpha$ and $\beta$ are nuisance parameters to be determined.

4. Determining the fiducial absolute magnitude $M_B^{\text{SN}}$ by combining data from multiple calibrator SnIa:

   \begin{equation}
       M_B^{\text{SN}} = m_B^{\text{corr}} - \mu_{\text{cal}}
       \label{eq:sniacal}
   \end{equation}
 where $\mu_{\text{cal}}$ is the distance modulus determined from the calibrated first-rung indicators.

The goal of this second rung is to precisely determine $M_B^{\text{SN}}$, $\alpha$, and $\beta$, which together allow SnIa to be used as standardizable candles for cosmological distance measurements.

The accuracy of the SN Ia calibration is critical for the determination of the Hubble constant. Recent analyses, such as the SH0ES project \cite{Riess:2021jrx}, have focused on refining this calibration to achieve percent-level precision in $H_0$ measurements. However, a fundamental assumption underlying this calibration process warrants careful consideration:

A crucial premise in the cosmic distance ladder methodology is that the calibration parameters remain constant across all rungs. This assumption implies that the physics governing the behavior of distance indicators (Cepheids, TRGB, SnIa) does not change significantly from local to cosmological scales. This assumption could be violated\cite{Perivolaropoulos:2022khd,Wojtak:2022bct} under several scenarios:

\begin{itemize}
    \item \textbf{Environmental Changes:} Systematic differences in the environments of distance indicators between rungs (e.g., dust properties, metallicity gradients) could affect their observed properties.
    
    \item \textbf{Evolutionary Effects:} Changes in stellar populations or galaxy properties over cosmic time might alter the characteristics of distance indicators.
    
    \item \textbf{Fundamental Physics Transitions:} A change in fundamental physics at some distance or time in the recent cosmological past could affect the behavior of distance indicators.
\end{itemize}

\paragraph{Distances, Redshifts, and Past Times of Each Rung:}
To contextualize these considerations, it's important to understand the typical distances, redshifts, and corresponding past times associated with each rung:

\begin{enumerate}
    \item \textbf{First Rung (Local Calibrators):}
        \begin{itemize}
            \item Distances: $\sim 10$ pc to $\sim 10$ Mpc
            \item Redshifts: $z \lesssim 0.002$
            \item Past time: $\sim 0 - 30$ Myr ago
            \item Examples: Milky Way Cepheids, LMC, NGC 4258
        \end{itemize}
    
    \item \textbf{Second Rung (SnIa Calibration):}
        \begin{itemize}
            \item Distances: $\sim 10$ Mpc to $\sim 40$ Mpc
            \item Redshifts: $0.002 \lesssim z \lesssim 0.01$
            \item Past time: $\sim 30 - 130$ Myr ago
            \item Examples: Galaxies hosting both Cepheids/TRGB/Miras/JAGB and SnIa
        \end{itemize}
    
    \item \textbf{Third Rung (Hubble Flow SnIa):}
        \begin{itemize}
            \item Distances: $\gtrsim 40$ Mpc
            \item Redshifts: $0.01 \lesssim z \lesssim 0.15$
            \item Past time: $\sim 130$ Myr - $1.8$ Gyr ago
            \item Note: Upper limit chosen to minimize cosmic acceleration effects
        \end{itemize}
\end{enumerate}

The transition between these rungs spans several orders of magnitude in distance and covers a significant portion of recent cosmic history ($\sim 1.8$ Gyr). This range extends from the present day to when the Universe was about 87\% of its current age. The past time estimates are based on a standard $\Lambda$CDM cosmology with $H_0 \approx 70$ km s$^{-1}$ Mpc$^{-1}$ and $\Omega_m \approx 0.3$.

This temporal span underscores the importance of validating the assumption of calibration consistency across these scales. Over this time, subtle changes in stellar populations, galactic environments, and even potentially in fundamental physics could occur, potentially affecting our distance indicators.

Potential violations of the calibration consistency assumption across these timescales could introduce systematic biases in $H_0$ measurements. Therefore, it is crucial to carefully consider and test for any time-dependent effects that might influence our distance indicators and ultimately our determination of the Hubble constant.

\subsubsection{Third Rung: Extending to Cosmological Scales}

The third rung of the cosmic distance ladder extends measurements to cosmological scales, primarily utilizing Type Ia supernovae (SnIa) as standardizable candles. While the calibration process of SnIa using Cepheids has been discussed in previous subsections, here we describe how these calibrated SnIa are employed to determine the Hubble constant.

\paragraph{Cosmographic Expansion Approach}
The determination of $H_0$ from SnIa data involves a cosmographic expansion of the Hubble parameter $H(z)$. This expansion is typically carried out to first order \cite{Visser:2004bf,Cattoen:2007sk}:

\begin{equation}
    H(z) = H_0 [1 + (1+q_0)z + \mathcal{O}(z^2)]
    \label{eq:Hz}
\end{equation}

where $q_0$ is the deceleration parameter. The luminosity distance $d_L$ can then be expressed as:

\begin{equation}
    d_L(z) = \frac{c}{H_0} [z + \frac{1}{2}(1-q_0)z^2 + \mathcal{O}(z^3)]
    \label{eq:dl}
\end{equation}

For low redshifts ($z \lesssim 0.1$), we can use this cosmographic expansion approach to measure $H_0$ using SnIa as standardizable candles. This method minimizes the dependence on specific cosmological models while still capturing the essence of cosmic expansion \cite{Riess:2016jrr,Riess:2019cxk}.

\paragraph{Distance Modulus and Fitting Procedure}
We start with the distance modulus equation:

\begin{equation}
m_\text{th}(z) = M_B + 5 \log_{10}\left[\frac{d_L(z)}{\text{Mpc}}\right] + 25
\label{eq:mth}
\end{equation}

where $m_\text{th}(z)$ is the theoretical apparent magnitude, $M_B$ is the absolute magnitude, and $d_L(z)$ is the luminosity distance. Using equation \eqref{eq:dl} in \eqref{eq:mth}, we obtain:

\begin{equation}
\begin{split}
m_\text{th}(z) = & M_B + 5 \log_{10}\left(cz \left[1 + \frac{1}{2}(1-q_0)z\right]\right) \\
& + 5 \log_{10}\left(\frac{c/H_0}{1 \text{ Mpc}}\right) + 25
\end{split}
\label{eq:mth_expanded}
\end{equation}

where the observables are $m_\text{th}$ and $z$, and the parameters are fit from SnIa data.

To break the degeneracy between $M_B$ and $H_0$, we use local measurements ($z < 0.01$, typically within 40 Mpc) of $M_B$ using relative distance indicators like Cepheid variables or the Tip of the Red Giant Branch (TRGB) \cite{Freedman:2019jwv}. We then assume this locally calibrated $M_B$ applies to the SnIa in the Hubble flow ($0.01 < z < 0.15$).

The fitting procedure involves minimizing:
\begin{equation}
\chi^2(H_0, q_0) = \sum_i \frac{[m_{\text{obs},i} - m_\text{th}(z_i; H_0, q_0)]^2}{\sigma_i^2}
\label{eq:chi2}
\end{equation}
where $m_{\text{obs},i}$ are the observed magnitudes and $\sigma_i$ are the associated uncertainties.

\paragraph{Practical Approaches}
In practice, two main approaches are employed:

\begin{enumerate}
    \item \textbf{Fixed $q_0$ Approach:} The deceleration parameter $q_0$ is fixed to its value derived from a fiducial cosmological model (e.g., Planck 2018 best-fit $\Lambda$CDM), and $H_0$ is determined solely from the SnIa data \cite{Riess:2019cxk,Riess:2020fzl}.
    
    \item \textbf{Joint Fit Approach:} Both $H_0$ and $q_0$ are simultaneously fit using the SnIa data from the third rung \cite{Camarena:2019moy,Camarena:2021jlr}. This approach reduces model dependence but may increase statistical uncertainties.
\end{enumerate}

The choice between these approaches can impact the resulting $H_0$ value and its associated uncertainties, highlighting the importance of careful consideration of the underlying cosmological model assumptions \cite{Efstathiou:2021ocp}.

\paragraph{Systematic Uncertainties}
Several sources of systematic uncertainty must be carefully addressed in this process:

\begin{itemize}
    \item \textbf{SnIa Standardization:} The intrinsic scatter in SnIa absolute magnitudes after standardization contributes to the uncertainty in $H_0$ \cite{Scolnic:2017caz}.
    
    \item \textbf{Local Calibration:} Uncertainties in the local calibration of $M_B$ using Cepheids or TRGB propagate directly to the $H_0$ measurement \cite{Freedman:2019jwv,Riess:2020fzl}.
    
    \item \textbf{Redshift Cut:} The choice of the upper redshift limit for the Hubble flow sample can affect the results due to increasing cosmological model dependence at higher redshifts \cite{Riess:2016jrr}.
    
    \item \textbf{Peculiar Velocities:} Corrections for peculiar velocities, especially important at lower redshifts, can introduce uncertainties \cite{Davis:2011hv,Dhawan:2020xmp}.
\end{itemize}

Recent analyses have made significant progress in addressing these systematics, leading to more robust $H_0$ measurements \cite{Riess:2021jrx,Freedman:2021ahq}. However, the persistent tension between local $H_0$ measurements and those inferred from early universe observations underscores the need for continued scrutiny and refinement of these methods \cite{DiValentino:2021izs,Perivolaropoulos:2021jda}.

For a recent pedagogical and detailed description of the distance ladder method, including advanced techniques and potential systematic effects, see Ref. \cite{Perivolaropoulos:2022khd}.

\paragraph{Complementary Secondary Indicators}
While SnIa serve as the primary tool for $H_0$ determination in the third rung, several other secondary distance indicators provide valuable cross-checks and complementary information:

\begin{itemize}
    \item \textbf{Surface Brightness Fluctuations (SBF):} This method \cite{Tonry:2000aa} utilizes the statistical properties of pixel-to-pixel variations in galaxy images to estimate distances.
    
    \item \textbf{L-$\sigma$ Relation for HII galaxies:} This technique \cite{Chavez:2014ria} leverages the correlation between HII region luminosity and gas velocity dispersion.
    
    \item \textbf{Tully-Fisher Relation:} This empirical relation \cite{Haridasu:2024ask,Russell:2008it,Sandage:2006ik} between galaxy rotational velocity and luminosity provides an independent distance measure for spiral galaxies.
\end{itemize}

These complementary methods serve to validate and refine the distance scale established by SnIa, enhancing the robustness of $H_0$ measurements.

\subsection{One Step Methods}
One step methods for measuring the Hubble constant, \(H_0\), provide an alternative to the traditional distance ladder approach\cite{Verde:2019ivm}. Unlike the multi-tiered distance ladder that relies on a series of interconnected rungs, one step methods seek to estimate cosmological distances directly from a single distance indicator without the need for intermediate calibrators. This section provides an overview of many of these methods, focusing on those that are independent from the sound horizon scale standard ruler and describing their unique advantages and limitations. One step methods are less susceptible to the cumulative errors and biases that can arise from local astrophysical effects, such as variations in metallicity, extinction, or stellar population characteristics. However, it is important to recognize that while one step methods are less dependent on local astrophysics, they generally offer less precision compared to the well-established distance ladder techniques, primarily due to the inherent challenges in measuring cosmological distances directly from singular observations. In the following subsections we discuss the main one step methods in some more detail.

\subsubsection{Sunyaev-Zel'dovich Effect}

The Sunyaev-Zel'dovich (SZ) effect, first predicted by Rashid Sunyaev and Yakov Zel'dovich in 1969, has emerged as a powerful tool in observational cosmology \cite{Carlstrom:2002na,Birkinshaw:1998qp}. This effect, which arises from the interaction between cosmic microwave background (CMB) photons and high-energy electrons in galaxy clusters, provides a method for measuring cosmological distances that is independent of the traditional cosmic distance ladder.

\paragraph{Physical Basis}
The SZ effect is primarily composed of two components:

\begin{itemize}
    \item \textbf{Thermal SZ effect:} The dominant component, resulting from the inverse Compton scattering of CMB photons by hot electrons in the intracluster medium (ICM).
    \item \textbf{Kinetic SZ effect:} A subdominant component caused by the bulk motion of the electron gas relative to the CMB rest frame.
\end{itemize}

This subsection focuses on the thermal SZ effect due to its greater magnitude and cosmological utility.

\paragraph{Mathematical Formulation}
The thermal SZ effect manifests as a frequency-dependent distortion of the CMB spectrum. The change in the CMB intensity due to the SZ effect is given by:

\begin{equation}
    \frac{\Delta I_{\nu}}{I_0} = y \cdot g(x)
    \label{eq:sz_intensity}
\end{equation}

where:
\begin{itemize}
    \item $I_0 = \frac{2(k_B T_{\text{CMB}})^3}{(hc)^2}$ is the undistorted CMB intensity
    \item $y$ is the Comptonization parameter
    \item $g(x)$ is the spectral function
    \item $x = \frac{h\nu}{k_B T_{\text{CMB}}}$ is the dimensionless frequency
\end{itemize}

The Comptonization parameter $y$ represents the integrated electron pressure along the line of sight:

\begin{equation}
    y = \int \frac{k_B T_e}{m_e c^2} n_e \sigma_T dl
    \label{eq:comptonization}
\end{equation}
where $T_e$ is the electron temperature, $n_e$ is the electron number density, $\sigma_T$ is the Thomson cross-section, and the integral is along the line of sight.

The spectral function $g(x)$ is given by:
\begin{equation}
    g(x) = \frac{x^4 e^x}{(e^x - 1)^2} \left( x \frac{e^x + 1}{e^x - 1} - 4 \right) \left( 1 + \delta_{\text{SZ}}(x, T_e) \right)
    \label{eq:spectral_function}
\end{equation}
where $\delta_{\text{SZ}}(x, T_e)$ accounts for relativistic corrections, which become significant for high-temperature clusters ($k_B T_e \gtrsim 10$ keV).

\paragraph{Observational Characteristics}
The SZ effect has several unique properties that make it valuable for cosmological studies\cite{Ade:2015fva,Hasselfield:2013wf,Bocquet:2018ukq,Salvati:2017rsn}:

\begin{itemize}
    \item It is independent of redshift, allowing for the detection of high-redshift clusters.
    \item The SZ surface brightness is proportional to the integrated pressure of the ICM, providing a direct probe of cluster thermodynamics.
    \item The effect causes a decrease in CMB intensity at frequencies below $\sim$218 GHz and an increase above this frequency, creating a distinctive spectral signature.
\end{itemize}

\paragraph{Application to Cosmology}
One of the most significant applications of the SZ effect is in measuring the Hubble constant ($H_0$). This method combines SZ observations with X-ray measurements of galaxy clusters to derive distances. The process involves:

\begin{enumerate}
    \item SZ observations provide a measure of $\int n_e T_e dl$.
    \item X-ray observations provide a measure of $\int n_e^2 \Lambda(T_e) dl$, where $\Lambda(T_e)$ is the X-ray cooling function.
    \item Assuming a geometry for the cluster (often a spherical isothermal $\beta$-model), these observations can be combined to solve for the angular diameter distance $D_A$.
    \item The Hubble constant is then derived using the relation $H_0 = cz/D_A$ for low-redshift clusters.
\end{enumerate}

\paragraph{Recent Results}
A notable study by Reese et al. \cite{Reese:2003ya} used this technique with a sample of 41 galaxy clusters to derive:

\begin{equation}
    H_0 = 61 \pm 3 \text{ (stat.)} \pm 18 \text{ (syst.)} \text{ km s}^{-1} \text{ Mpc}^{-1}
    \label{eq:h0_result}
\end{equation}

where the uncertainties are given at 68\% confidence. The large systematic uncertainty reflects challenges in modeling cluster geometry and evolution.

\paragraph{Future Prospects}
Ongoing and future SZ surveys, such as those conducted with the Atacama Cosmology Telescope (ACT) and the South Pole Telescope (SPT), promise to significantly improve the precision of SZ-based cosmological measurements. These improvements will come from:

\begin{itemize}
    \item Larger cluster samples reducing statistical uncertainties
    \item Better understanding of cluster physics and improved modeling techniques
    \item Combination with other cosmological probes to break degeneracies and reduce systematic uncertainties
\end{itemize}

The SZ effect, with its unique redshift-independence and direct probe of cluster physics, continues to be a valuable tool in modern cosmology, complementing other methods in our quest to understand the universe's expansion history and large-scale structure.

\subsubsection{Megamasers as Standard Rulers}

Megamasers, particularly those associated with water molecules in active galactic nuclei (AGN), have emerged as powerful standard rulers for measuring extragalactic distances \cite{Reid:2012hm,Pesce:2020xfe}. These astrophysical phenomena provide a direct geometric method for determining the Hubble constant, $H_0$, independent of the cosmic distance ladder.

\paragraph{Physical Basis}
Water megamasers are typically found in the accretion disks surrounding supermassive black holes in AGN. The 22 GHz ($\lambda = 1.35 cm$)  emission line of water results from the 6$_{16}$ → 5$_{23}$ rotational transition. This maser emission occurs under specific conditions:

\begin{itemize}
    \item Temperatures of $\sim$300-1000 K
    \item High water molecule densities ($n_{H_2O} \sim 10^8 - 10^{10}$ cm$^{-3}$)
    \item Presence of a pumping mechanism (likely collisional)
\end{itemize}

The bright, compact nature of these masers, combined with their Keplerian motion in the accretion disk, makes them ideal targets for very long baseline interferometry (VLBI) observations.

\paragraph{Observational Technique}
VLBI observations of megamasers provide two key measurements:

\begin{enumerate}
    \item Precise spatial positions of individual maser spots
    \item Doppler shifts of maser lines, yielding line-of-sight velocities
\end{enumerate}

These observations typically reveal a characteristic pattern in position-velocity space, with systemic masers near the center of the disk and high-velocity masers on the disk's edge, tracing a Keplerian rotation curve.

\paragraph{Physical Model and Distance Determination}
The geometry and dynamics of the maser-emitting accretion disk allow for the application of Kepler's laws of motion. The enclosed mass $M$ (dominated by the central supermassive black hole) can be derived from the rotation speed $v$ and radius $r$ of the maser spots:

\begin{equation}
    v^2 = \frac{GM}{r}
    \label{eq:kepler}
\end{equation}

where $G$ is the gravitational constant. VLBI observations provide both $v$ (from Doppler shifts) and the angular separation $\theta$ between maser spots.

The distance $D$ to the galaxy can then be calculated using:

\begin{equation}
    D = \frac{v}{\theta \omega}
    \label{eq:distance}
\end{equation}
where $\omega$ is the angular velocity of the maser spots. This equation combines the physical size of the disk (derived from equation \eqref{eq:kepler}) with its angular size as observed from Earth.

The Hubble constant can then be estimated using:

\begin{equation}
    H_0 = \frac{v_H}{D}
    \label{eq:hubble}
\end{equation}
where $v_H$ is the Hubble flow velocity, typically derived from the galaxy's redshift after correcting for peculiar motions.

\paragraph{Advantages and Challenges}
The megamaser method offers several advantages:

\begin{itemize}
    \item Direct geometric distance measurement
    \item Independence from the cosmic distance ladder
    \item High precision (potentially $\sim$3\% per galaxy)
    \item Applicability to galaxies at cosmologically significant distances
\end{itemize}

However, challenges include:

\begin{itemize}
    \item Rarity of suitable megamaser systems
    \item Complexity of accretion disk models (e.g., warping, non-circular motions)
    \item Need for high-sensitivity VLBI observations
\end{itemize}

\paragraph{Recent Results}
The Megamaser Cosmology Project (MCP) has been at the forefront of using this technique. A notable example is their study of NGC 5765b, an Sa-b galaxy hosting water megamasers \cite{Gao:2015tqd}. Key findings include:

\begin{itemize}
    \item Confirmation of a thin, sub-parsec Keplerian disk
    \item Evidence for a spiral density wave influencing accretion dynamics
    \item Angular-diameter distance to NGC 5765b: $126.3 \pm 11.6$ Mpc
    \item Hubble constant estimate: $H_0 = 66.0 \pm 6.0$ km s$^{-1}$ Mpc$^{-1}$
\end{itemize}

This measurement incorporated secular drifts of maser features and a detailed model of the disk's warped structure to refine the $H_0$ estimate.

\paragraph{Future Prospects}
The megamaser method continues to be refined and applied to new systems. Future developments may include:

\begin{itemize}
    \item Increased sample size through more sensitive surveys
    \item Improved modeling of non-Keplerian motions and disk structure
    \item Combination with other distance measurement techniques to constrain systematic errors
\end{itemize}

As an independent probe of $H_0$, megamasers play a crucial role in addressing the tension between early and late universe measurements of the expansion rate. Their continued study promises to contribute significantly to our understanding of cosmology and the dynamics of galactic nuclei.

\subsubsection{Strong Gravitational Lensing Time Delays}

Strong gravitational lensing time delays have emerged as a powerful, independent method for measuring the Hubble constant ($H_0$) \cite{Birrer:2022chj,H0LiCOW:2016tzl,Birrer:2020tax,Millon:2019slk,Dobler:2013rda,Baxter:2020qlr}. This technique leverages the phenomenon of gravitational lensing, where a massive foreground object (typically a galaxy or galaxy cluster) bends light from a background source (often a quasar), creating multiple images. The difference in light travel time between these images provides a direct probe of cosmological distances.

\paragraph{Theoretical Foundation}

The fundamental principle of time-delay cosmography is rooted in general relativity. Light from a background source, when lensed by a foreground mass, travels along different paths with varying lengths and gravitational potentials. This results in the lensed images arriving at the observer at different times. These time delays, when accurately measured, are directly related to the geometry of the Universe and the mass distribution of the lens, thus providing a means to constrain cosmological parameters, particularly $H_0$.

The key quantity in time-delay cosmography is the time-delay distance $D_{\Delta t}$, which is a combination of angular diameter distances:

\begin{equation}
    D_{\Delta t} \equiv (1+z_d) \frac{D_d D_s}{D_{ds}},
    \label{eq:time_delay_distance}
\end{equation}

where $z_d$ is the redshift of the deflector (lens), $D_d$ is the angular diameter distance to the deflector, $D_s$ is the angular diameter distance to the source, and $D_{ds}$ is the angular diameter distance from the deflector to the source.

The observed time delay $\Delta t_{AB}$ between two lensed images A and B is related to the Fermat potential difference $\Delta \phi_{AB}$ by:

\begin{equation}
    \Delta t_{AB} = \frac{D_{\Delta t}}{c} \Delta \phi_{AB},
    \label{eq:time_delay}
\end{equation}

where $c$ is the speed of light. The Fermat potential $\phi(\boldsymbol{\theta})$ at a position $\boldsymbol{\theta}$ on the sky is defined as:

\begin{equation}
    \phi(\boldsymbol{\theta}) = \frac{(\boldsymbol{\theta} - \boldsymbol{\beta})^2}{2} - \psi(\boldsymbol{\theta}),
    \label{eq:fermat_potential}
\end{equation}

with $\boldsymbol{\beta}$ being the source position and $\psi(\boldsymbol{\theta})$ the lensing potential.

\paragraph{Methodology and Challenges}

The Time-Delay COSMOgraphy (TDCOSMO) collaboration, building upon earlier efforts such as COSMOGRAIL, H0LiCOW, and STRIDES, has refined the methodology for using strong lensing time delays to measure $H_0$ \cite{Millon:2019slk,Birrer:2020tax}. Their approach involves:

\begin{enumerate}
    \item Precise time delay measurements through long-term, high-cadence monitoring of lensed quasars.
    \item Advanced lens mass modeling, incorporating both parametric and non-parametric approaches.
    \item Rigorous accounting for line-of-sight effects using spectroscopic data and numerical simulations.
    \item A comprehensive Bayesian framework for inferring cosmological parameters.
\end{enumerate}

Key challenges in this method include:

\begin{itemize}
    \item Accurate characterization of the lens mass distribution, including the effects of dark matter.
    \item Proper accounting for the mass-sheet degeneracy and other potential degeneracies in lens modeling.
    \item Quantifying and mitigating the impact of line-of-sight structures on the lensing potential.
    \item Obtaining sufficiently long and well-sampled light curves for precise time delay measurements.
\end{itemize}

\paragraph{Current Results and Implications}

The landscape of $H_0$ measurements from time-delay cosmography has evolved significantly in recent years:

\begin{itemize}
    \item TDCOSMO's early combined analysis of seven lens systems yielded $H_0 = 74.2 \pm 1.6$ km s$^{-1}$ Mpc$^{-1}$ (2.2\% precision) \cite{Wong:2019kwg}, aligning with distance ladder results.
    
    \item Du et al. (2023) \cite{Du:2023zsz} found $H_0 = 71.5^{+4.4}_{-3.0}$ km s$^{-1}$ Mpc$^{-1}$ using time-delay galaxy lenses combined with gamma-ray bursts.
    
    \item Birrer et al. (2020) \cite{Birrer:2020tax}, in a joint analysis of TDCOSMO and SLACS samples, obtained $H_0 = 67.4^{+4.1}_{-3.2}$ km s$^{-1}$ Mpc$^{-1}$, shifting towards lower values more consistent with Planck results.
    
    \item Most recently, TDCOSMO (2024) \cite{TDCOSMO:2024rwr} reported $H_0 = 65^{+23}_{-14}$ km s$^{-1}$ Mpc$^{-1}$ from time-delay cosmography of WGD 2038-4008, albeit with large uncertainties.
\end{itemize}

This progression suggests that as methods are refined and datasets expanded, time-delay cosmography and related techniques are increasingly yielding $H_0$ measurements more in line with Planck CMB-based estimates, rather than supporting the higher values from local distance ladder methods.

\paragraph{Future Prospects}

The future of time-delay cosmography is promising, with several developments on the horizon:

\begin{itemize}
    \item Upcoming wide-field surveys like the Vera C. Rubin Observatory's Legacy Survey of Space and Time (LSST) are expected to discover thousands of new lensed quasars, dramatically increasing the sample size.
    \item Advancements in adaptive optics and space-based imaging will improve the resolution and quality of lens imaging, enhancing mass modeling capabilities.
    \item Improved spectroscopic follow-up will better constrain line-of-sight effects and lens dynamics.
    \item Integration with other cosmological probes, such as strongly lensed supernovae and gravitational waves, may provide complementary constraints and help break degeneracies.
\end{itemize}

As TDCOSMO and other collaborations continue to refine their methodology, expand their lens samples, and explore potential systematic effects, time-delay cosmography is poised to play an increasingly important role in precision cosmology and in addressing the Hubble tension. The method's strength lies in its direct approach, utilizing fundamental physics without reliance on distance ladder calibrations, making it a valuable complement to other cosmological probes.

\subsubsection{Horizon at Matter-Radiation Equality as a Standard Ruler}

The horizon at matter-radiation equality presents an alternative standard ruler for cosmological distance measurements, distinct from the widely used sound horizon scale \cite{Philcox:2022sgj,Brieden:2022heh,DAmico:2020ods}. This section elucidates the principles underlying this approach, presents key equations, and discusses recent research findings.

\paragraph{Theoretical Foundation}

The horizon at matter-radiation equality, characterized by the wavenumber $k_{\text{eq}}$, represents the comoving scale entering the horizon when the energy densities of matter and radiation are equal. This scale significantly influences the shape of the matter power spectrum, particularly affecting the turnover point, which is crucial for determining the broadband shape of the linear power spectrum at scales $k \sim k_{\text{eq}}$.

The comoving wavenumber at matter-radiation equality, $k_{\text{eq}}$, is given by:

\begin{equation}
    k_{\text{eq}} = \sqrt{2 \Omega_m H_0^2 a_{\text{eq}}^{-1}},
    \label{eq:k_eq}
\end{equation}
where $\Omega_m$ is the matter density parameter, $H_0$ is the Hubble constant, and $a_{\text{eq}}$ is the scale factor at matter-radiation equality. This can be rewritten in terms of the redshift at matter-radiation equality, $z_{\text{eq}}$, as:
\begin{equation}
    k_{\text{eq}} = \sqrt{2 \Omega_m H_0^2 (1 + z_{\text{eq}})}.
    \label{eq:k_eq_z}
\end{equation}

The scale $k_{\text{eq}}$ sets the peak of the matter power spectrum and is crucial for determining the linear power spectrum's shape at scales around and larger than $k_{\text{eq}}$.

\paragraph{Methodology and Applications}

Recent studies, such as those by Philcox et al. (2022) \cite{Philcox:2022sgj}, Brieden et al. (2022) \cite{Brieden:2022heh}, and D'Amico et al. (2020) \cite{DAmico:2020ods}, have utilized $k_{\text{eq}}$ to derive independent constraints on the Hubble constant, $H_0$, within a $\Lambda$CDM framework. This approach combines measurements from various cosmological probes:

\begin{itemize}
    \item Galaxy surveys: Providing information on the matter power spectrum shape
    \item CMB lensing: Offering complementary constraints on matter clustering
    \item Supernovae: Constraining the expansion history of the Universe
\end{itemize}

The key advantage of this method is its independence from the sound horizon scale, which is typically used in BAO studies. This independence makes it less sensitive to early universe physics and potential new physics that might affect the pre-recombination era.

\paragraph{Recent Results}

Philcox et al. (2022) \cite{Philcox:2022sgj} combined data from BOSS galaxy power spectra, Planck CMB lensing, and the Pantheon+ supernova compilation to derive an $H_0$ value of:
\begin{equation}
    H_0 = 64.8^{+2.2}_{-2.5} \text{ km s}^{-1}\text{Mpc}^{-1} \quad (68\% \text{ confidence level}).
    \label{eq:H0_result}
\end{equation}

This result is consistent with Planck CMB-based estimates and in tension with local distance ladder measurements, providing an independent cross-check on the Hubble tension.

\paragraph{Advantages and Limitations}

The use of the horizon at matter-radiation equality as a standard ruler offers several advantages:

\begin{itemize}
    \item Independence from early universe physics: Less sensitive to potential new physics affecting the pre-recombination era.
    \item Complementarity: Provides a check on results from BAO and CMB analyses.
    \item Robustness: Relies on well-understood physics of the matter-radiation transition.
\end{itemize}

However, there are also limitations to consider:

\begin{itemize}
    \item Precision: Currently less precise than BAO measurements.
    \item Model dependence: Assumes a $\Lambda$CDM cosmology.
    \item Systematic uncertainties: Requires careful modeling of nonlinear effects in galaxy clustering.
\end{itemize}

\paragraph{Future Prospects}

The potential of this method is likely to grow with forthcoming galaxy surveys such as DESI, Euclid, and the Vera C. Rubin Observatory's LSST. These surveys will provide more precise measurements of the matter power spectrum over a wider range of scales and redshifts, potentially improving constraints on $k_{\text{eq}}$ and, consequently, on $H_0$.

Furthermore, combining this approach with other probes, such as gravitational lensing and the kinetic Sunyaev-Zel'dovich effect, could break degeneracies and further improve constraints on cosmological parameters.

In conclusion, using the horizon at matter-radiation equality as a standard ruler provides a compelling alternative for cosmological distance measurements. Its robustness against potential new early universe physics makes it a valuable tool for modern cosmology, offering an independent check on the consistency of cosmological measurements and theories. As observational data improve and analysis techniques are refined, this method is poised to play an increasingly important role in addressing key cosmological questions, including the Hubble tension.

\subsubsection{Cosmic Chronometers}

Cosmic chronometers offer a novel approach to measure the expansion rate of the Universe, $H(z)$, directly from observational data without relying on the standard cosmological model \cite{Moresco:2023zys,Gomez-Valent:2018hwc,Moresco:2016nqq,Moresco:2016mzx}. This method, based on the differential age approach, utilizes the time difference ($\Delta t$) between two distinct epochs of the Universe to infer the Hubble parameter. This section outlines the foundational principles, key equations, and practical applications of this method.

\paragraph{Theoretical Foundation}

The cosmic chronometer approach leverages the age of the oldest galaxies at different redshifts to estimate the rate of cosmic expansion. The core assumption is that the most massive, passively evolving galaxies, often termed 'red and dead' galaxies, form their stellar populations early and evolve with minimal subsequent star formation. These galaxies serve as 'chronometers' because their age can be closely associated with the age of the Universe at their formation redshift.

\paragraph{Methodology}

The Hubble parameter, $H(z)$, which quantifies the expansion rate of the Universe, can be directly related to the observable quantities of redshift ($z$) and time ($t$) via the differential age method. The fundamental relation is given by:

\begin{equation}
    H(z) = -\frac{1}{1+z} \frac{dz}{dt}.
    \label{eq:H_z_differential}
\end{equation}

In practice, this is approximated using finite differences:

\begin{equation}
    H(z) \approx -\frac{1}{1+z} \frac{\Delta z}{\Delta t},
    \label{eq:H_z_discrete}
\end{equation}
where $\Delta z$ is the redshift difference between two galaxy populations, and $\Delta t$ is the differential age between these populations, estimated from their stellar populations.

\paragraph{Implementation and Challenges}

The implementation of the cosmic chronometer method involves several key steps:

\begin{enumerate}
    \item Selection of passively evolving galaxies: Identifying massive, early-type galaxies with minimal ongoing star formation.
    \item Spectroscopic observations: Obtaining high-quality spectra to determine redshifts and estimate stellar population ages.
    \item Age determination: Using stellar population synthesis models to estimate the age of the galaxies based on their spectral features.
    \item Differential age calculation: Computing the age difference between galaxy populations at different redshifts.
    \item Hubble parameter estimation: Applying Equation \eqref{eq:H_z_discrete} to calculate $H(z)$ at various redshifts.
\end{enumerate}

This method faces several challenges:
\begin{itemize}
    \item Accurate age determination: Requires precise stellar population models and high-quality spectroscopic data.
    \item Sample selection: Ensuring a consistent selection of passively evolving galaxies across cosmic time.
    \item Systematics: Addressing potential biases in age estimates and selection effects.
    \item Limited redshift range: Currently restricted to $z \lesssim 2$ due to observational limitations.
\end{itemize}

\paragraph{Illustrative Example}

Consider two populations of galaxies at redshifts $z_1$ and $z_2$, with $z_2 > z_1$. Spectroscopic studies determine their ages to be $t_1$ and $t_2$ respectively. Let $\Delta t = t_2 - t_1 = 1$ Gyr and $\Delta z = z_2 - z_1 = 0.1$. The Hubble parameter at the mean redshift $\bar{z} = (z_1 + z_2)/2$ can be calculated as:

\begin{equation}
    H(\bar{z}) = -\frac{1}{1 + \bar{z}} \frac{\Delta z}{\Delta t}.
    \label{eq:H_z_example}
\end{equation}

Assuming $\bar{z} = 1.5$ and converting units appropriately:

\begin{equation}
    H(1.5) \approx -\frac{1}{2.5} \frac{0.1}{1 \text{ Gyr}} \approx 78 \text{ km s}^{-1} \text{ Mpc}^{-1},
    \label{eq:H_z_result}
\end{equation}

where the negative sign indicates expansion.

\paragraph{Recent Results and Future Prospects}

Recent studies have applied the cosmic chronometer method to constrain $H(z)$ over a wide redshift range. For instance, Moresco et al. (2016) \cite{Moresco:2016mzx} provided measurements of $H(z)$ up to $z \sim 2$, finding good agreement with $\Lambda$CDM predictions but with larger uncertainties at high redshifts.

Future prospects for this method include:
\begin{itemize}
    \item Improved stellar population models: Enhancing the accuracy of age determinations.
    \item Larger galaxy samples: Upcoming surveys like DESI and Euclid will provide larger samples of passive galaxies.
    \item Extended redshift range: Near-infrared spectroscopy may allow measurements at $z > 2$.
    \item Synergies with other probes: Combining with BAO and SNe Ia data for tighter cosmological constraints.
\end{itemize}

The cosmic chronometer method offers a unique, model-independent approach to measure the expansion history of the Universe. While it currently provides less precise measurements compared to other probes, its independence from the cosmic distance ladder and standard cosmological assumptions makes it a valuable tool for cross-validating results from other methods and potentially uncovering new physics beyond the standard model.

\subsubsection{Standard Sirens and Kilonovae}

Standard sirens, a term coined for gravitational waves (GWs) emitted from compact binary mergers, provide a unique method for measuring the Hubble constant, $H_0$, directly from the gravitational wave signal without requiring a cosmic distance ladder \cite{Abbott2017Nature,LIGOScientific:2017adf,Hotokezaka:2018dfi,Chen:2017rfc,Ando:2012hna,Bulla:2022ppy}. This section details the principles behind using GWs and their electromagnetic counterparts, kilonovae, as cosmological probes, highlighting their potential to resolve the Hubble tension.

\paragraph{Gravitational Waves as Standard Sirens}

Gravitational waves emitted from mergers of binary neutron stars (BNS) or neutron star-black hole (NS-BH) binaries offer a direct measure of the luminosity distance to the source\cite{Abbott2017Nature}. The amplitude of the gravitational wave signal is inversely proportional to this distance, and by combining it with the redshift of the host galaxy, $H_0$ can be directly inferred.

The relationship between the strain amplitude $h$ of the gravitational wave and the luminosity distance $D_L$ is given by the quadrupole formula:

\begin{equation}
    h = \frac{4(G\mathcal{M})^{5/3}}{c^4D_L} (\pi f)^{2/3} \mathcal{F}(\text{angle, orientation}),
    \label{eq:gw_amplitude}
\end{equation}
where $G$ is the gravitational constant, $c$ is the speed of light, $\mathcal{M}$ is the chirp mass, $f$ is the frequency of the gravitational wave, and $\mathcal{F}$ is a function of the binary's inclination angle and sky position.

\paragraph{Kilonovae as Electromagnetic Counterparts}

A kilonova (KN) is a thermal electromagnetic emission resulting from the radioactive decay of heavy elements synthesized in the neutron-rich ejecta of a neutron star merger. This emission is quasi-isotropic and can be seen from a wide range of viewing angles, making it an excellent counterpart for observing gravitational wave events. The combined observation of a gravitational wave and a kilonova from the same event provides a powerful tool for cosmology, as it adds an independent measurement of the electromagnetic flux, which aids in breaking degeneracies in the gravitational wave signal.

\paragraph{Hubble Constant Measurement}

The relationship between luminosity distance $D_L$ and redshift $z$ in the local universe is expressed by:

\begin{equation}
    v_H = cz = H_0 D_L,
    \label{eq:hubble_law}
\end{equation}

where $v_H$ is the Hubble flow velocity. For nearby sources where peculiar velocities are significant, this relation is modified to:

\begin{equation}
    v_H = cz_{\text{helio}} - v_{\text{pec,LOS}} = H_0 D_L,
    \label{eq:hubble_law_corrected}
\end{equation}
where $z_{\text{helio}}$ is the heliocentric redshift and $v_{\text{pec,LOS}}$ is the line-of-sight peculiar velocity.

\paragraph{The GW170817 Event}

The landmark event GW170817, where both gravitational waves and a kilonova \cite{LIGOScientific:2017vwq,Abbott2017Nature} were observed, serves as a prime example of this method. The gravitational wave signal provided a measurement of the luminosity distance, while the identification of the host galaxy NGC 4993 led to a redshift measurement. Combining these measurements allowed for an independent estimate of $H_0$.

\paragraph{Breaking the Inclination Degeneracy}

The inclination of the binary system affects the amplitude of the gravitational wave signal, introducing a degeneracy between inclination and distance. The kilonova emission, being anisotropic and dependent on the viewing angle, provides additional constraints on the system's orientation. By combining the gravitational wave data with kilonova light curves and spectra, the inclination-distance degeneracy can be partially broken, leading to more precise $H_0$ measurements.

\paragraph{Statistical Approach}

For events without identified electromagnetic counterparts, a statistical approach can be employed. This method involves cross-correlating the GW-inferred sky localizations with galaxy catalogs to obtain a probabilistic estimate of the host galaxy and its redshift. While less precise than the direct counterpart method, this approach allows for the use of a larger sample of GW events.

\paragraph{Current Results and Future Prospects}

Recent analyses combining multiple GW events have provided competitive constraints on $H_0$. For example, the LIGO-Virgo-KAGRA collaboration reported:

\begin{equation}
    H_0 = 68^{+8}_{-7} \text{ km s}^{-1} \text{ Mpc}^{-1} \quad (68\% \text{ credible interval}),
    \label{eq:H0_result}
\end{equation}
based on a sample of binary black hole mergers and GW170817 \cite{LIGOScientific:2021aug}.

Future prospects for this method are promising:

\begin{itemize}
    \item Increased event rate: Advanced detectors will observe more BNS and NS-BH mergers, improving statistical precision.
    \item Improved detector sensitivity: Enhanced strain sensitivity will lead to more precise distance measurements.
    \item Better kilonova models: Improved theoretical understanding and observations of kilonovae will enhance constraints on system parameters.
    \item Synergies with other probes: Combining standard siren measurements with other cosmological probes can provide powerful constraints on cosmological parameters.
\end{itemize}

The combination of gravitational waves as standard sirens and kilonovae offers a promising route to independently measure the Hubble constant. These multi-messenger observations not only provide a direct measurement of cosmic distances but also enhance our understanding of the universe's expansion through the precise calibration of $H_0$ without reliance on traditional distance ladders. As the field of gravitational wave astronomy matures, standard sirens are poised to play a crucial role in resolving the Hubble tension and advancing our understanding of cosmology.

\subsubsection{Gamma-Ray Attenuation}

Gamma-ray attenuation through interaction with the extragalactic background light (EBL) provides a unique method to study the universe's expansion and to estimate cosmological parameters such as the Hubble constant ($H_0$) \cite{Dominguez:2019jqc,Dominguez:2023rxa,Desai:2019pfl}. This subsection explores the principle behind gamma-ray attenuation, its theoretical framework, and its application in deriving cosmological parameters.

\paragraph{Theoretical Framework}

High-energy gamma-ray photons emitted by distant astrophysical sources such as blazars and gamma-ray bursts interact with the EBL—comprising photons mainly in the ultraviolet, optical, and infrared wavelengths—resulting in electron-positron pair production \cite{Dominguez:2010wxv,Desai:2019pfl}. This process attenuates the gamma-ray flux from these sources, and the extent of attenuation depends on the gamma-ray photon energy, the density of the EBL, and the distance the gamma rays have traveled, which is linked to the redshift of the source.

The optical depth $\tau$ for gamma rays traveling from a source at redshift $z$ with observed energy $E$ can be described by the equation:

\begin{eqnarray}
    \tau(E, z) &=& \int_0^z \frac{dl}{dz'} dz' \int_{-1}^{1} d\mu \frac{1-\mu}{2} \nonumber \\
    && \times \int_{\epsilon_{\text{th}}}^\infty d\epsilon \, n(\epsilon, z') \nonumber \\
    && \times \sigma_{\gamma\gamma}(E(1+z'), \epsilon, \mu)
    \label{eq:optical_depth}
\end{eqnarray}

where:
\begin{itemize}
    \item $\frac{dl}{dz'}$ is the cosmology-dependent line-of-sight element,
    \item $\mu = \cos\theta$, with $\theta$ being the interaction angle,
    \item $\epsilon$ is the energy of the EBL photon,
    \item $\epsilon_{\text{th}}$ is the threshold energy for pair production,
    \item $n(\epsilon, z')$ is the proper number density of EBL photons,
    \item $\sigma_{\gamma\gamma}$ is the pair production cross-section.
\end{itemize}

The pair production cross-section $\sigma_{\gamma\gamma}$ is given by:

\begin{equation}
    \sigma_{\gamma\gamma}(\beta) = \frac{3\sigma_T}{16}(1-\beta^2) \left[ (3-\beta^4)\ln\frac{1+\beta}{1-\beta} - 2\beta(2-\beta^2) \right],
    \label{eq:cross_section}
\end{equation}

where $\sigma_T$ is the Thomson cross-section and $\beta = \sqrt{1-\frac{2m_e^2c^4}{E\epsilon(1-\mu)}}$ is the velocity of the electron/positron in the center-of-mass frame.

The line-of-sight element is given by:
\begin{equation}
    \frac{dl}{dz} = \frac{c}{H_0} \frac{1}{(1+z)\sqrt{\Omega_m(1+z)^3 + \Omega_\Lambda}},   \label{eq:line_of_sight}
\end{equation}
where $c$ is the speed of light, $H_0$ is the Hubble constant, $\Omega_m$ is the matter density parameter, and $\Omega_\Lambda$ is the dark energy density parameter.

\paragraph{Cosmic Gamma-Ray Horizon (CGRH)}

The cosmic gamma-ray horizon (CGRH) defines the energy at which the universe becomes opaque to gamma rays due to EBL attenuation ($\tau = 1$). By measuring the CGRH as a function of redshift, one can directly probe the EBL and thus indirectly measure the expansion rate of the universe. The CGRH is defined implicitly by:

\begin{equation}
    \tau(E_{\text{CGRH}}, z) = 1,
    \label{eq:cgrh}
\end{equation}
where $E_{\text{CGRH}}$ is the energy of the CGRH at redshift $z$.

\paragraph{Application to Cosmology}

Using improved EBL models from comprehensive galaxy surveys such as CANDELS \cite{Grogin:2011ua} and gamma-ray attenuation data from telescopes like Fermi-LAT, recent studies \cite{Dominguez:2019jqc,Dominguez:2023rxa,Desai:2019pfl} have derived new optical depths and used these to calculate the CGRH. This approach provides an one-step sound horizon free estimate for $H_0$.

The methodology typically involves:

\begin{enumerate}
    \item Constructing an EBL model based on galaxy survey data.
    \item Calculating gamma-ray optical depths using Eq. \eqref{eq:optical_depth}.
    \item Determining the CGRH from Eq. \eqref{eq:cgrh}.
    \item Fitting cosmological parameters to the observed CGRH.
\end{enumerate}

Recent results from this method have yielded\cite{Dominguez:2023rxa}:
\begin{equation}
    H_0 = 62.4^{+4.1}_{-3.9} \text{ km s}^{-1} \text{ Mpc}^{-1} \quad (\Omega_m = 0.32 \text{ fixed}),
    \label{eq:H0_result1}
\end{equation}
\begin{equation}
    H_0 = 65.1^{+6.0}_{-4.9} \text{ km s}^{-1} \text{ Mpc}^{-1} \quad (\Omega_m = 0.19 \pm 0.08),
    \label{eq:H0_result2}
\end{equation}
showcasing the utility of this method in cosmology.

\paragraph{Strengths and Limitations}

The gamma-ray attenuation method offers several advantages:
\begin{itemize}
    \item Independence from traditional distance ladders.
    \item Probes the universe at high redshifts ($z \sim 1-5$).
    \item Sensitive to the EBL, providing insights into galaxy evolution.
    \item Offers a crucial cross-check in the era of precision cosmology.
\end{itemize}

However, it also faces challenges:
\begin{itemize}
    \item Requires accurate modeling of the EBL.
    \item Sensitive to assumptions about blazar spectra and their evolution.
    \item Limited by the current statistics of high-energy gamma-ray observations.
\end{itemize}

\paragraph{Future Prospects}

The future of this method looks promising with:
\begin{itemize}
    \item Upcoming gamma-ray telescopes like the Cherenkov Telescope Array (CTA) providing better statistics and higher energy reach.
    \item Improved EBL models from future galaxy surveys.
    \item Advancements in our understanding of blazar physics and evolution.
\end{itemize}

In conclusion, gamma-ray attenuation provides a powerful cosmological tool that leverages high-energy astrophysics to probe the fundamental parameters governing the universe's expansion. As our understanding of the EBL and gamma-ray sources improves, this method is poised to play an increasingly important role in resolving cosmological tensions and advancing our knowledge of the universe.

\subsubsection{BAO Measurement of $H_0$ Independent of Sound Horizon Drag Scale}

Baryon Acoustic Oscillations (BAO) have long served as a powerful standard ruler in cosmology. Traditionally, BAO measurements depend on the sound horizon scale at the drag epoch, $r_d$. However, recent developments have enabled measurements of the Hubble constant ($H_0$) using BAO data that are independent of $r_d$, offering new insights into cosmological parameters and potentially resolving tensions in modern cosmology \cite{Pogosian:2024ykm,Liu:2024txl,Gomez-Valent:2021hda}.

\paragraph{Theoretical Framework}

The key to this approach lies in the BAO observable $\beta_\perp(z)$, which represents the ratio of the comoving angular diameter distance to the sound horizon scale:

\begin{equation}
    \beta_\perp(z) = \frac{D_M(z)}{r_d},
    \label{eq:beta_perp}
\end{equation}
where $D_M(z)$ is the comoving angular diameter distance at redshift $z$. In a flat $\Lambda$CDM cosmology, neglecting radiation density at the redshifts of interest, $\beta_\perp(z)$ can be expressed as:
\begin{equation}
    \beta_\perp(z) = \frac{c}{r_d H_0} \int_0^z \frac{dz'}{\sqrt{\Omega_m(1+z')^3 + 1 - \Omega_m}},
    \label{eq:beta_perp_integral}
\end{equation}
where $c$ is the speed of light, $H_0$ is the Hubble constant, and $\Omega_m$ is the matter density parameter.
\paragraph{Methodology}
The innovative aspect of this method involves several key steps:
\begin{enumerate}
    \item Measure $\beta_\perp(z)$ at multiple redshifts.
    \item Analyze the shape of $\beta_\perp(z)$ as a function of redshift.
    \item Introduce a prior on $\Omega_m h^2$ from CMB measurements.
\end{enumerate}
Measuring $\beta_\perp(z)$ at multiple redshifts allows us to constrain two key parameters: $r_d h$ and $\Omega_m$. This is because the shape of $\beta_\perp(z)$ as a function of redshift depends on $\Omega_m$, while its overall normalization is set by $r_d h$, where $h = H_0/(100 \text{ km s}^{-1} \text{Mpc}^{-1})$.

However, there exists a degeneracy between $r_d$ and $h$ that cannot be broken by BAO measurements alone. To resolve this, an additional constraint is introduced: a prior on $\Omega_m h^2$ from CMB measurements. This prior is particularly useful because it's derived from the overall shape of the CMB power spectrum and is largely independent of assumptions about the sound horizon scale.

With these three pieces of information:
\begin{itemize}
    \item $\beta_\perp(z)$ measurements at multiple redshifts (constraining $r_d h$ and $\Omega_m$)
    \item The shape of $\beta_\perp(z)$ (further constraining $\Omega_m$)
    \item A prior on $\Omega_m h^2$ from CMB data
\end{itemize}
There is now sufficient information to solve for three parameters: $r_d$, $\Omega_m$, and $h$ (or equivalently, $H_0$).

\paragraph{Mathematical Framework}

The method can be formalized as follows:

\begin{enumerate}
    \item From BAO measurements, we obtain $\beta_\perp(z_i)$ at various redshifts $z_i$.
    \item We fit these measurements to the theoretical model given by Eq. \eqref{eq:beta_perp_integral}, which depends on $r_d h$ and $\Omega_m$.
    \item We incorporate the CMB prior on $\Omega_m h^2$, which we can write as:
    \begin{equation}
        \Omega_m h^2 = (\Omega_m h^2)_\text{CMB} \pm \sigma_{\Omega_m h^2},
        \label{eq:cmb_prior}
    \end{equation}
    where $(\Omega_m h^2)_\text{CMB}$ is the central value from CMB measurements and $\sigma_{\Omega_m h^2}$ is its uncertainty.
    \item We can then solve for $h$ using:
    \begin{equation}
        h = \sqrt{\frac{(\Omega_m h^2)_\text{CMB}}{\Omega_m}},
        \label{eq:h_solution}
    \end{equation}
    where $\Omega_m$ is determined from the fit to $\beta_\perp(z)$.
    \item Finally, we can determine $r_d$ using the best-fit value of $r_d h$ and the derived value of $h$.
\end{enumerate}

\paragraph{Results and Implications}

Using data from the Dark Energy Spectroscopic Instrument (DESI) alongside Planck CMB measurements, this method yields \cite{Pogosian:2024ykm}:
\begin{equation}
    H_0 = 69.88 \pm 0.93 \text{ km s}^{-1}\text{Mpc}^{-1}.
    \label{eq:H0_result}
\end{equation}
This result is particularly noteworthy as it falls between the lower values typically obtained from CMB analyses and the higher values from local distance ladder measurements, potentially offering a path towards resolving the Hubble tension.

\paragraph{Strengths and Limitations}

The strength of this approach lies in its ability to provide an $H_0$ estimate that is independent of sound horizon-based calibrations. Key advantages include:

\begin{itemize}
    \item Independence from early-universe physics assumptions that affect $r_d$.
    \item Utilization of well-understood BAO physics.
    \item Combination of late-time (BAO) and early-time (CMB) information.
\end{itemize}

However, limitations and potential sources of systematic error should be considered:

\begin{itemize}
    \item Dependence on the accuracy of BAO measurements across multiple redshifts.
    \item Reliance on the $\Lambda$CDM framework for the $\beta_\perp(z)$ model.
    \item Potential sensitivity to the CMB prior on $\Omega_m h^2$.
\end{itemize}

\paragraph{Future Prospects}

This methodology not only serves as a consistency check for $\Lambda$CDM but also opens up new avenues for understanding early universe physics and potentially resolving cosmological tensions. Future prospects include:

\begin{itemize}
    \item Improved BAO measurements from upcoming surveys like DESI, Euclid, and the Vera C. Rubin Observatory.
    \item Refinement of CMB priors from future CMB experiments.
    \item Extension to models beyond $\Lambda$CDM to test for new physics.
\end{itemize}

In conclusion, this $r_d$-independent method for measuring $H_0$ using BAO data represents a significant advance in cosmological analysis. By decoupling the determination of $H_0$ from assumptions about the sound horizon scale, it provides a valuable new tool for probing both the early and late-time universe, potentially shedding light on the Hubble tension by providing a one step method for measuring $H_0$ that is independent of the sound horizon scale.

\section{Statistical analysis of grouped \(H_0\) measurements}

In this section, we aim to construct a comprehensive compilation of the most available measurements of the Hubble constant (\(H_0\)) focusing mainly in the last 5 years. These measurements will be categorized into two distinct groups: those based on the distance ladder and those derived from one-step methods that are independent of the distance ladder. The primary goal is to analyze and compare the statistical properties of each group, focusing on the mean \(H_0\) and its standard deviation. This comparison will help assess the consistency between the two groups and their alignment with the sound horizon-based measurement from the Planck 2018 data. By doing so, we aim to gain insights into the potential sources of the Hubble tension and explore whether discrepancies arise from systematic effects or fundamental differences in measurement techniques. 

\subsection{Presentation of measurements}

In this subsection, we present a detailed overview of the \(H_0\) measurements compiled into two categories: distance ladder-based measurements and one-step, distance ladder-independent measurements. We begin with the distance ladder-based measurements, which rely on multiple calibration steps to estimate cosmic distances.

\subsubsection{Distance Ladder-Based Measurements}

In this subsection we briefly describe the $H_0$ measurements of Table \ref{tab:distance_ladder_h0_measurements} based on distance ladder measurements used in the present analysis 

\begin{table*}[ht]
\centering
\begin{tabular}{>{\ttfamily}l l l l}
\toprule
\textbf{Index} & \textbf{First Author (Year)} & \textbf{Measured \( H_0 \) (km/s/Mpc)} & \textbf{Method} \\
\midrule
1 & \href{https://arxiv.org/abs/1908.10883}{Huang (2019)} \cite{Huang:2019yhh} & \( 73.3 \pm 4.0 \) & Mira calibrators \\
2 & \href{https://arxiv.org/abs/2004.14499}{Kourkchi (2020)} \cite{Kourkchi:2020iyz} & \( 76.0 \pm 3.4 \) & Tully Fisher + Cepheid + TRGB \\
3 & \href{https://arxiv.org/abs/2101.02221}{Blakeslee (2021)} \cite{Blakeslee:2021rqi} & \( 73.3 \pm 3.1 \) & SBF + Cepheids + TRGB \\
4 & \href{https://arxiv.org/abs/2106.15656}{Freedman (2021,24)} \cite{Freedman:2021ahq,Freedman:2024eph} & \( 69.8 \pm 2.3 \) & TRGB calibrators \\
5 & \href{https://arxiv.org/abs/2112.04510}{Riess (2021)} \cite{Riess:2021jrx} & \( 73.04 \pm 1.04 \) & Cepheid (SH0ES) \\
6 & \href{https://arxiv.org/abs/2203.04241}{Dhawan (2022)} \cite{Dhawan:2022yws} & \( 76.94 \pm 6.4 \) & TRGB calibrators \\
7 & \href{https://arxiv.org/abs/2211.07657}{Dhawan (2022)} \cite{Dhawan:2022gac} & \( 74.82 \pm 1.81 \) & BayeSN + Cepheids \\
8 & \href{https://arxiv.org/abs/2211.07657}{Dhawan (2022)} \cite{Dhawan:2022gac} & \( 70.92 \pm 2.63 \) & BayeSN + TRGB \\
9 & \href{https://arxiv.org/pdf/2204.10866}{Kenworthy (2022)} \cite{Kenworthy:2022jdh} & \( 73.1 \pm 2.5 \) & two rung distance ladder \\
10 & \href{https://arxiv.org/abs/2304.06693}{Scolnic (2023)} \cite{Scolnic:2023mrv} & \( 73.22 \pm 2.06 \) & TRGB calibrators (SH0ES) \\
11 & \href{https://arxiv.org/pdf/2308.01875}{Uddin (2023)} \cite{Uddin:2023iob} & \( 71.76 \pm 1.32 \) & TRGB + Cepheids with SnIa and SBF (B band) \\
12 & \href{https://arxiv.org/pdf/2308.01875}{Uddin (2023)} \cite{Uddin:2023iob} & \( 73.22 \pm 1.45 \) & TRGB + Cepheids with SnIa and SBF (H Band) \\
13 & \href{https://arxiv.org/abs/2305.17243}{de Jaeger (2023)} \cite{deJaeger:2023vkm} & \( 74.1 \pm 8 \) & SnII Weighted mean SCM \\
14 & \href{https://arxiv.org/abs/2312.08423}{Huang (2023)} \cite{Huang:2023frr} & \( 72.37 \pm 2.97 \) & Mira calibrators \\
15 & \href{https://arxiv.org/abs/2404.16261}{Chavez (2024)} \cite{Chavez:2024twa} & \( 73.1 \pm 2.3 \) & $L -\sigma$ HII galaxies \\
16 & \href{https://arxiv.org/pdf/2401.04777}{Li (2024)} \cite{Li:2024yoe,Li:2024pjo} & \( 74.7 \pm 3.1 \) & JAGB calibartors with SnIa \\
17 & \href{https://arxiv.org/pdf/2408.03474}{Lee (2024)} \cite{Lee:2024qzr} & \( 68.00 \pm 2.7 \) & JAGB calibartors with SnIa \\
18 & \href{https://arxiv.org/pdf/2408.06153}{Freedman (2024)} \cite{Freedman:2024eph} & \( 72.05 \pm 3.6 \) & Cepheid calibrators with SnIa \\
19 & \href{https://arxiv.org/pdf/2408.03660}{Boubel (2024)} \cite{Boubel:2024cqw} & \( 73.3 \pm 4.08 \) & Tully Fisher + Cepheid + TRGB \\
19 & \href{https://arxiv.org/pdf/2408.03660}{Boubel (2024)} \cite{Boubel:2024cqw} & \( 73.3 \pm 4.08 \) & Tully Fisher + Cepheid + TRGB \\
20 & \href{https://arxiv.org/pdf/2408.13842}{Said (2024)} \cite{Said:2024pwm} & \( 76.05 \pm 4.90 \) & DESI FP relation + SBF \\
\bottomrule
\end{tabular}
\caption{Distance ladder-dependent measurements of \( H_0 \) with corresponding first authors and methods.}
\label{tab:distance_ladder_h0_measurements}
\end{table*}

1. Mira Calibrators \cite{Huang:2019yhh}:In Ref. \cite{Huang:2019yhh}, Mira variables\cite{Whitelock:2008mz,Huang:2018byc} are utilized to calibrate the luminosity of Type Ia supernovae (SnIa) and subsequently measure the Hubble constant (\(H_0\)). The analysis involves a year-long observation of O-rich Miras in the galaxy NGC 1559, host to SN Ia 2005df, using the Hubble Space Telescope's WFC3/IR. Mira variables, identifiable by their Period-Luminosity Relation (PLR) in the near-infrared, are calibrated using the geometric distance to NGC 4258, known from water megamaser measurements, and the Large Magellanic Cloud (LMC) via detached eclipsing binaries. The study reports a distance modulus for NGC 1559 of \(\mu_{1559} = 31.41 \pm 0.05\) (statistical) \(\pm 0.06\) (systematic) mag, leading to an \(H_0\) measurement of \(73.3 \pm 4.0\) km s\(^{-1}\) Mpc\(^{-1}\) when combined with calibrations from the NGC 4258 megamaser and the LMC. This approach positions Miras not only as a complementary method to Cepheids for calibrating SnIa but also as a potential tool for resolving discrepancies in measurements of \(H_0\) from the early and late Universe, contributing to ongoing discussions about the "Hubble tension."

2. Tully Fisher + Cepheid + TRGB \cite{Kourkchi:2020iyz}: In the study by Kourkchi et al. (2020) \cite{Kourkchi:2020iyz}, the Tully-Fisher relation, which correlates the rotational velocity of spiral galaxies with their luminosity, is finely calibrated using both Cepheid variables and the Tip of the Red Giant Branch (TRGB) as part of the Cosmicflows-4 project. This calibration effort is extensive, involving measurements across optical (SDSS u, g, r, i, z) and infrared (WISE W1 and W2) bands for a large sample of spiral galaxies. The study incorporates a subsample of approximately 600 spiral galaxies located in 20 galaxy clusters to refine the Tully-Fisher relation. The calibration process includes determining inclinations through an online graphical interface and utilizing new HI linewidth data primarily from the Arecibo Legacy Fast ALFA Survey. By integrating these calibrations, the study achieves a preliminary determination of the Hubble constant, \( H_0 = 76.0 \pm 1.1 \) (statistical) \( \pm 2.3 \) (systematic) km/s/Mpc, offering a crucial stepping stone in resolving discrepancies in the measurement of the universe's expansion rate. This comprehensive approach underscores the intricate multi-tier calibration necessary in the distance ladder methodology, aiming to reduce the scatter in the Tully-Fisher relation and hence improve the accuracy of distance measurements.

3. SBF + Cepheids + TRGB: In the study by Blakeslee et al. (2021) \cite{Blakeslee:2021rqi}, the Surface Brightness Fluctuations (SBF) method is employed in conjunction with Cepheid variables and the Tip of the Red Giant Branch (TRGB) to calibrate and measure distances to galaxies. Specifically, the SBF technique, which utilizes the variance in brightness between different sections of a galaxy to infer its distance, is applied to infrared observations from the Hubble Space Telescope's Wide Field Camera 3. This approach is complemented by calibrations using Cepheids and TRGB, which serve as standard candles. These calibrations are refined with the latest parallax data from Gaia EDR3 and direct geometric measurements from masers in galaxies like NGC 4258. Through this multifaceted calibration strategy, the study achieves a highly precise measurement of the Hubble constant, \( H_0 = 73.3 \pm 0.7 \) (statistical) \( \pm 2.4 \) (systematic) km/s/Mpc. This value is consistent with other high-precision measurements using Cepheids and Type Ia supernovae, reinforcing the robustness of the combined SBF, Cepheid, and TRGB approach within the framework of the cosmic distance ladder.

4. TRGB Calibrators \cite{Freedman:2021ahq,Freedman:2024eph} (see also 18.): In the study by Freedman et al. (2021), the Tip of the Red Giant Branch (TRGB) method is utilized to calibrate the Hubble constant (\(H_0\)) with a focus on reducing systematic uncertainties inherent in cosmic distance measurements. This paper consolidates various TRGB calibrations, which are demonstrated to be internally consistent at the 1\% level, with additional verification from Gaia Early Data Release 3 (EDR3) providing a lower 5\% accuracy level due to Gaia's angular covariance bias. The TRGB-calibrated distance to a sample of Type Ia supernovae from the Carnegie Supernova Project yields an \(H_0\) value of \(69.8 \pm 0.6 \, (\text{stat}) \pm 1.6 \, (\text{sys})\) km/s/Mpc. This value aligns closely with Cosmic Microwave Background (CMB) measurements under the standard \( \Lambda \)CDM model, suggesting no significant tension with early universe estimates, thereby indicating that the observed discrepancies in \(H_0\) might stem from unrecognized systematic errors rather than new physics. This study underscores the efficacy of the TRGB method, noted for its straightforward underlying physics and minimal systematic susceptibilities, in refining the local measurement of \(H_0\).

5. Cepheid (SH0ES) \cite{Riess:2021jrx}: In the comprehensive study by Riess et al. (2022), the SH0ES team employed Cepheid variables as primary distance calibrators within a refined distance ladder framework to determine the local value of the Hubble constant (\(H_0\)). Utilizing the Hubble Space Telescope (HST), they observed Cepheid variables in 42 Type Ia supernova (SN Ia) host galaxies, significantly enhancing the sample size and thereby refining the precision of \(H_0\). These Cepheids were geometrically calibrated using parallaxes from Gaia EDR3, masers in NGC 4258, and detached eclipsing binaries in the Large Magellanic Cloud, ensuring uniform measurement across different systems to mitigate zeropoint errors. The resultant measurement from the Cepheid-SN Ia calibration yielded an \(H_0\) value of \(73.04 \pm 1.04\) km/s/Mpc, incorporating systematic uncertainties and demonstrating a significant tension with the Planck CMB observations under the \(\Lambda\)CDM model. This tension underscores potential discrepancies in cosmological models or the presence of unaccounted systematic errors in distance measurement techniques.

6. TRGB Calibrators \cite{Dhawan:2022yws}: In the study presented by Dhawan et al. (2022), a uniform distance ladder is constructed using Type Ia supernovae (SnIa) observed by the Zwicky Transient Facility (ZTF), with absolute calibration based on the Tip of the Red Giant Branch (TRGB) method. The TRGB calibration leverages the core helium flash luminosity of low-mass stars at the end of the Red Giant Branch, providing a standard candle that is less sensitive to environmental conditions compared to Cepheids. This method enables probing SN Ia host galaxies of all types within a volume-limited sample, thereby minimizing systematic errors related to host-galaxy bias. In their pilot study, Dhawan et al. apply this approach to SN 2021rhu in NGC 7814, using high-cadence ZTF observations and HST data for TRGB distance estimation. The TRGB-calibrated distance to NGC 7814 leads to an \(H_0\) measurement of \(76.94 \pm 6.4\) km/s/Mpc, reflecting significant potential for resolving the Hubble tension if the TRGB calibration can be applied across a broader sample of SnIa, supported by future JWST observations.

7-8. BayeSN + Cepheids or TRGB \cite{Dhawan:2022gac}: In the study by Dhawan et al. (2022), the Hubble constant (\(H_0\)) is estimated using a multi-faceted approach combining Bayesian analysis of Type Ia supernovae (SnIa) light curves across optical and near-infrared (NIR) spectrums with the BayeSN model, alongside distance calibrations based on Cepheid variables and the Tip of the Red Giant Branch (TRGB). This hierarchical Bayesian model, BayeSN, integrates data across a wide wavelength range to provide a continuous spectral energy distribution, enhancing the accuracy of distance measurements by utilizing the intrinsic properties of SnIa, which are less scattered in NIR. Two separate calibrations are employed: one using Cepheid distances to 37 host galaxies of 41 SnIa, yielding an estimated \(H_0 = 74.82 \pm 0.97 \, (\text{stat}) \pm 0.84 \, (\text{sys})\) km/s/Mpc, and another using TRGB distances to 15 host galaxies of 18 SnIa, resulting in \(H_0 = 70.92 \pm 1.14 \, (\text{stat}) \pm 1.49 \, (\text{sys})\) km/s/Mpc. These estimates underscore the sensitivity of \(H_0\) measurements to the choice of distance ladder rung, with significant differences arising from the methodologies and inherent systematics of the Cepheid and TRGB calibrations. The analysis demonstrates the potential of combining optical and NIR data to refine local measurements of \(H_0\) and thereby address the existing tension between local and early-universe estimates.

9. Kenworthy (2022) \cite{Kenworthy:2022jdh}: This analysis presents a measurement of the Hubble constant using a two-rung distance ladder consisting of Cepheid variables and redshifts of their host galaxies. The authors report $H_0 = 73.1^{+2.6}_{-2.3}$ km s$^{-1}$ Mpc$^{-1}$. The method employs 35 Cepheid host galaxies from the SH0ES team at $z \leq 0.011$, using their distances and redshifts to measure the Hubble flow. To mitigate the significant impact of peculiar velocities at these low redshifts, the analysis incorporates detailed modeling of the local velocity field using two independent reconstructions. The authors implement a hierarchical Bayesian model to account for selection effects, covariances in peculiar velocities, and systematic uncertainties in the Cepheid distances. Multiple analysis variants are considered to assess the impact of different assumptions about sample selection and peculiar velocity corrections. The final result is derived by combining four model variants, yielding a 3.5\% precision measurement that is in 2.6$\sigma$ tension with Planck. While not as precise as the full three-rung SH0ES ladder, this measurement provides an important cross-check that is independent of potential systematic uncertainties in Type Ia supernovae.

10. TRGB Calibrators (SH0ES)\cite{Scolnic:2023mrv}: In this analysis, the Tip of the Red Giant Branch (TRGB) method is leveraged within the SH0ES project framework to refine the measurement of the Hubble constant (\(H_0\)). The TRGB, a luminous standard candle used for constructing distance ladders, is calibrated using an unsupervised algorithm, Comparative Analysis of TRGBs (CATs), which minimizes variance among multiple halo fields across various host galaxies without reliance on subjective adjustments. This study applies CATs to an expanded sample of SN Ia hosts standardized to the geometric anchor NGC 4258, alongside the Pantheon+ SN Ia sample. By standardizing the apparent TRGB tips across different hosts using empirical correlations between TRGB measurements and contrast ratios, the study achieves a refined measurement of \(H_0 = 73.22 \pm 2.06\) km/s/Mpc. This enhanced approach addresses discrepancies in earlier TRGB studies by incorporating systematic algorithmic adjustments that account for variations in the TRGB measurement process, thus contributing to a more consistent and accurate cosmological distance scale.

11-12. Uddin (2023) \cite{Uddin:2023iob}: This analysis presents measurements of the Hubble constant using Type Ia supernovae (SnIa) from the Carnegie Supernova Project I and II, calibrated with three independent distance indicators: Cepheid variables, the tip of the red giant branch (TRGB), and surface brightness fluctuations (SBF). The authors report $H_0 = 71.76 \pm 0.58 \text{ (stat)} \pm 1.19 \text{ (sys)}$ km s$^{-1}$ Mpc$^{-1}$ from B-band data and $H_0 = 73.22 \pm 0.68 \text{ (stat)} \pm 1.28 \text{ (sys)}$ km s$^{-1}$ Mpc$^{-1}$ from H-band data. The analysis employs a Bayesian framework to fit a warped disk model to the SN Ia data, incorporating peculiar velocity corrections and marginalizing over nuisance parameters. Systematic uncertainties are derived by combining results from different calibrators. The methodology critically examines various sources of uncertainty, including sample selection effects, host galaxy properties, and the wavelength dependence of SN Ia standardization. While the precision is comparable to other recent $H_0$ measurements, the use of multiple calibrators provides insight into systematic differences between distance scales. The authors find that dust may not be the primary driver of the host-mass step in SN Ia luminosities, contrary to some previous studies.

13. SnII + Cepheids + TRGB \cite{deJaeger:2023vkm}: In the study by de Jaeger and Galbany (2023) \cite{deJaeger:2023vkm}, Type II supernovae (SnII) are utilized as distance indicators. The paper employs both the Expanding Photosphere Method (EPM) and the Standard Candle Method (SCM) to derive distance estimates which are then used to calculate the Hubble constant (\(H_0\)). 

The EPM is a geometric approach that calculates distances by measuring the physical expansion of the supernova's photosphere and comparing it to its apparent brightness, while SCM employs a correlation between the supernova's luminosity at the plateau phase and its photospheric velocity. The SCM method is calibrated using Cepheids and TRGB, providing a link to the cosmic distance ladder. From the EPM, Ref. \cite{deJaeger:2023vkm} reports a range of \(H_0\) values, with the latest measurements giving \(H_0 = 84.70^{+2.28}_{-2.21}\) km/s/Mpc and \(H_0 = 75.57^{+2.04}_{-1.89}\) km/s/Mpc. The SCM, highlighted for its empirical approach, shows values such as \(H_0 = 75 \pm 7\) km/s/Mpc and \(H_0 = 75.8^{+5.2}_{-4.9}\) km/s/Mpc among others over the years (Table 1.1 of Ref. \cite{deJaeger:2023vkm}). 

In our analysis, we utilize the weighted mean of the SCM method measurements from Table 1.1 of \cite{deJaeger:2023vkm} Thus we use  \( 74.1 \pm 8 \). We use the average uncertainty of these measurements to account for their correlation.

14. Mira Calibrators \cite{Huang:2023frr}: In this analysis, Mira variables are employed to refine measurements of the Hubble constant (\(H_0\)), using a method that parallels the Cepheid distance ladder but incorporates Mira variables as an alternative or complementary standard candle. The analysis leverages a comprehensive dataset from the Hubble Space Telescope (HST), consisting of 11 epochs of F110W and 13 epochs of F160W observations, to identify and analyze 211 Mira variables within the galaxy M101. These variables, characterized by periods between 240 and 400 days, are used to establish a Period-Luminosity Relation (PLR), calibrated against geometric distances to the Large Magellanic Cloud (LMC) and the NGC 4258 galaxy, known for its water megamaser measurements. The calibration yields a distance modulus to M101 of \(\mu_{\text{M101}} = 29.10 \pm 0.06\) mag, aligning closely with values derived from other methods like Cepheids and TRGB. Subsequently, utilizing the calibrated Mira variables and the peak luminosity of Type Ia Supernova SN 2011fe, the study calculates \(H_0 = 72.37 \pm 2.97\) km s\(^{-1}\) Mpc\(^{-1}\), a precision of 4.1\%. This result not only supports the consistency of Mira variables within the cosmic distance ladder but also corroborates the higher \(H_0\) values observed in the local universe compared to those predicted from early universe measurements, contributing to the ongoing investigation into the "Hubble tension."

15. $L -\sigma$ HII Galaxies \cite{Chavez:2024twa}: In the study by Chávez et al. (2024), the $L-\sigma$ relation, which correlates the luminosity of Balmer lines with the velocity dispersion in HII galaxies, is employed as a cosmological distance indicator, extending to a redshift $z \approx 7.5$. This method capitalizes on the observation capabilities of the James Webb Space Telescope (JWST) to observe high-redshift HII galaxies, facilitating a comprehensive measurement of the Hubble constant ($H_0$) across a vast redshift range, covering 95\% of the Universe's history. Utilizing a dataset of 231 HII galaxies and extragalactic HII regions, the study applies Bayesian inference to refine cosmological parameters, yielding results for a flat Universe as $h = 0.731 \pm 0.039$, $\Omega_m = 0.302^{+0.12}_{-0.069}$, and $w_0 = -1.01^{+0.52}_{-0.29}$ (statistical uncertainties). These parameters are pivotal for understanding the expansion history of the Universe and are derived from a method that complements traditional distance indicators such as Cepheid variables and Type Ia supernovae within the cosmic distance ladder framework. The calibration of the $L-\sigma$ relation is cross-checked against Cepheid and Type Ia supernova measurements to ensure accuracy and consistency, enhancing the reliability of this method as a robust cross-check against other distance indicators. This approach not only underscores the uniformity of photo-kinematic properties of HII regions across cosmic time but also enhances our understanding of cosmic expansion dynamics, contributing significantly to the ongoing precision cosmology efforts.

16.  JAGB Calibrators (SH0ES)\cite{Li:2024yoe}: The recent analysis of Ref. \cite{Li:2024yoe} presents a study of JAGB stars using JWST NIRCam photometry in NGC 4258 and 4 hosts of 6 Type Ia supernovae (SN Ia): NGC 1448, NGC 1559, NGC 5584, and NGC 5643. Key findings include:

\begin{itemize}
\item JAGB clumps are readily apparent near $1.0 < F150W - F277W < 1.5$ and $m_{F150W} = 22-25$ mag
\item Various methods for assigning an apparent reference magnitude were tested, including mode, median, sigma-clipped mean, and modeled luminosity function parameter
\item Intra-host variations of up to $\sim$0.2 mag were found, significantly exceeding statistical uncertainties
\item The non-uniform shape of the JAGB luminosity function was observed, similar to that in the LMC and SMC
\item Broad agreement was found with distances measured from Cepheids, tip of the red giant branch (TRGB), and Miras
\item Different methods for estimating H0 yielded a range of 71 - 78 km s$^{-1}$ Mpc$^{-1}$
\item A fiducial result of $H_0 = 74.7 \pm 2.1(\text{stat}) \pm 2.3(\text{sys})$ km s$^{-1}$ Mpc$^{-1}$ was obtained
\end{itemize}

17.  JAGB Calibrators (Freedman) \cite{Lee:2024qzr}:
While the recent analysis by Li et al. (2024) \cite{Li:2024yoe} provides useful insights into the JAGB method using JWST NIRCam photometry, the more recent analysis by  Lee et al. (2024) \cite{Lee:2024qzr} presents a different approach and findings. 

Lee et al.  (2024) introduce a novel algorithm to identify the optimal location in a galaxy for applying the JAGB method, aiming to minimize crowding effects. This approach differs from Li et al. (2024), who examined intra-host variations across different regions of the galaxies.

\begin{itemize}
\item \textbf{Sample Size:} Lee et al. studied seven SN Ia host galaxies, while Li et al. examined NGC 4258 and 4 hosts of 6 SN Ia.
\item \textbf{Filters:} Lee et al. primarily used F115W (J-band equivalent) and F356W, with F444W for some galaxies. Li et al. used F150W and F277W.
\item \textbf{Analysis Approach:} Lee et al. employed a blind analysis, adding random offsets to the photometry that were only removed after finalizing the analysis. This blind approach was not mentioned in Li et al.'s study.
\end{itemize}

\paragraph{Photometry and Star Selection}

Both studies used DOLPHOT for photometry, but with some differences:

\begin{itemize}
\item Lee et al. used the warm-start mode in DOLPHOT, extracting photometry from F115W images first and then reducing LW images using the source list from the first run.
\item Lee et al. applied specific quality-metric cuts (Table 1 in their paper) to remove non-stellar sources, prioritizing sample purity over completeness.
\item Lee et al. calculated deprojected galactocentric radii for each source, using these to separate JAGB stars into 'inner' and 'outer' regions of each galaxy.
\end{itemize}

\paragraph{JAGB Identification and Measurement}
The methods for identifying and measuring JAGB stars differ between the two studies:

\begin{itemize}
\item Li et al. identified JAGB clumps near $1.0 < F150W - F277W < 1.5$ and $m_{F150W} = 22-25$ mag.
\item Lee et al. used F115W (J-band) for JAGB magnitude measurements, consistent with the theoretical basis of the JAGB method \cite{Madore:2020yqv}.
\item Lee et al. focused on the outer regions of galaxies to minimize crowding and reddening effects, whereas Li et al. examined intra-host variations across different regions.
\end{itemize}

\paragraph{Results and $H_0$ Measurements}

The two studies arrived at different H0 values:

\begin{equation}
H_0 (\text{Lee et al.}) = 67.96 \pm 1.85 (\text{stat}) \pm 1.90 (\text{sys}) \text{ km s}^{-1} \text{Mpc}^{-1}
\end{equation}

\begin{equation}
H_0 (\text{Li et al.}) = 74.7 \pm 2.1 (\text{stat}) \pm 2.3 (\text{sys}) \text{ km s}^{-1} \text{Mpc}^{-1}
\end{equation}

These results highlight a significant discrepancy, with Lee et al.'s measurement aligning more closely with CMB-based estimates, while Li et al.'s result is more consistent with Cepheid-based measurements.

18. Freedman (2024) \cite{Freedman:2024eph} (used also in 4.): This analysis presents measurements of the Hubble constant using three independent methods applied to JWST observations of 10 nearby galaxies hosting Type Ia supernovae: the Tip of the Red Giant Branch (TRGB), J-region Asymptotic Giant Branch (JAGB) stars, and Cepheid variables. The authors find $H_0 = 69.85 \pm 1.75$ (stat) $\pm 1.54$ (sys) km s$^{-1}$ Mpc$^{-1}$ for TRGB, $H_0 = 67.96 \pm 1.85$ (stat) $\pm 1.90$ (sys) km s$^{-1}$ Mpc$^{-1}$ for JAGB, and $H_0 = 72.05 \pm 1.86$ (stat) $\pm 3.10$ (sys) km s$^{-1}$ Mpc$^{-1}$ for Cepheids. Combining these methods and tying to SnIa yields $H_0 = 69.96 \pm 1.05$ (stat) $\pm 1.12$ (sys) km s$^{-1}$ Mpc$^{-1}$. The analysis employs a consistent calibration using NGC 4258 for all three methods, allowing for a direct comparison. The TRGB and JAGB methods show excellent agreement, while the Cepheid distances are systematically shorter. The authors critically assess various sources of uncertainty, including the small number of calibrating galaxies, and discuss the implications for the Hubble tension. They conclude that while their results are consistent with $\Lambda$CDM predictions, more data are needed to definitively resolve the tension. The study demonstrates the power of JWST for improving local distance measurements but also highlights remaining challenges in achieving 1

19. Boubel (2024) \cite{Boubel:2024cqw}: This analysis presents an improved method for measuring the Hubble constant ($H_0$) using the Tully-Fisher relation and a peculiar velocity model. The authors apply their technique to the Cosmicflows-4 catalog, simultaneously fitting the Tully-Fisher relation and peculiar velocity field for the full sample. They calibrate the zero-point using galaxies with independent distance measurements from Cepheids, TRGB, and SnIa. For the $i$-band sample, they find $H_0 = 73.3 \pm 2.1 \text{ (stat)} \pm 3.5 \text{ (sys)}$ km s$^{-1}$ Mpc$^{-1}$, while the $W1$-band yields $H_0 = 74.5 \pm 1.2 \text{ (stat)} \pm 2.6 \text{ (sys)}$ km s$^{-1}$ Mpc$^{-1}$. The method improves upon previous approaches by using the entire dataset to fit the Tully-Fisher relation, incorporating a comprehensive model with curvature and varying intrinsic scatter, and accounting for selection effects and Malmquist bias. However, the authors note that the precision is currently limited by systematic uncertainties in the absolute distance calibrators and the small number of galaxies with such calibrations. They critically assess various sources of uncertainty, including the choice of calibrators, peculiar velocity modeling, and galaxy inclination measurements. While the method shows promise for future large surveys, its full potential awaits resolution of ongoing debates regarding absolute distance scales.

20. Said et al. (2024) \cite{Said:2024pwm}: This analysis employed the Fundamental Plane (FP) relation for early-type galaxies in combination with Surface Brightness Fluctuation (SBF) measurements to estimate the Hubble constant. The study utilized data from the Dark Energy Spectroscopic Instrument (DESI) Peculiar Velocity Survey, focusing on a sample of early-type galaxies. The FP relation, which correlates the effective radius, velocity dispersion, and surface brightness of elliptical galaxies, was calibrated using SBF distances to nearby galaxies, particularly the Coma cluster. This approach yielded $H_0 = 76.05 \pm 4.90$ km s$^{-1}$ Mpc$^{-1}$. It's important to note that this method is dependent on the cosmic distance ladder and requires local calibration. The reliance on local calibrators, such as SBF measurements of the Coma cluster, means that the resulting $H_0$ value is sensitive to potential systematic errors in these calibrations. This dependence on local distance indicators places this measurement in the category of distance ladder dependent $H_0$ determinations, which often yield higher values compared to measurements that are independent of the distance ladder.

\subsubsection{One Step Distance Ladder-Independent Measurements}

In this subsection we briefly describe the $H_0$ measurements of Table \ref{tab:independent_h0_measurements} based on one step methods, independent of both distance ladder and sound horizon scale used in the present analysis

\begin{table*}[ht]
\centering
\begin{tabular}{>{\ttfamily}l l l l}
\toprule
\textbf{Index} & \textbf{First Author (Year)} & \textbf{Measured \( H_0 \) (km/s/Mpc)} & \textbf{Method} \\
\midrule
1 & \href{https://arxiv.org/abs/astro-ph/0306073}{Reese (2003)} \cite{Reese:2003ya} & \( 61 \pm 18 \) & SZ Effect \\
2 & \href{https://arxiv.org/abs/1207.7273}{Kuo (2013)} \cite{Kuo2013} & \( 68 \pm 9 \) & Megamasers (MCP) \\
3 & \href{https://arxiv.org/abs/1511.08311}{Gao (2015)} \cite{Gao:2015tqd} & \( 66.0 \pm 6.0 \) & Megamasers (MCP) \\
4 & \href{https://arxiv.org/abs/1912.08027}{Millon (2019)} \cite{Millon:2019slk} & \( 74.2 \pm 1.6 \) & Lensing TD TDCOSMO I. \\
5 & \href{https://arxiv.org/abs/2001.09213}{Pesce (2020)} \cite{Pesce:2020xfe} & \( 73.9 \pm 3.0 \) & Megamasers (MCP+SH0ES) \\
6 & \href{https://arxiv.org/abs/2007.02941}{Birrer (2020)} \cite{Birrer:2020tax} & \( 67.4 \pm 4 \) & Lensing TD TDCOSMO + SLACS \\
7 & \href{https://arxiv.org/abs/2111.03604}{Abbott (2021)} \cite{LIGOScientific:2021aug} & \( 68.0^{+8}_{-6} \) & 47 GW sources from GWTC-3 \\
8 & \href{https://arxiv.org/abs/2108.00581}{Wu (2021)} \cite{Wu:2021jyk} & \( 64.67^{+5.62}_{-4.66} \) & Fast Radio Bursts \\
9 & \href{https://arxiv.org/abs/2204.02984}{Philcox (2022)} \cite{Philcox:2022sgj} & \( 64.8 \pm 2.4 \) & \( T_{eq} \) standard ruler \\
10 & \href{https://arxiv.org/abs/2208.03960}{Zhang (2022)} \cite{Zhang:2022caa} & \( 65.9 \pm 3.0 \) & Cosmic Chronometers + HII galaxies \\
11 & \href{https://arxiv.org/abs/2205.09145}{Bulla (2022)} \cite{Bulla:2022ppy} & \( 69.6 \pm 5.5 \) & Gravitational Waves + Kilonovae \\
12 & \href{https://arxiv.org/abs/2307.09501}{Moresco (2023)} \cite{Moresco:2023zys} & \( 66.7 \pm 5.3 \) & Cosmic Chronometers \\
13 & \href{https://arxiv.org/abs/2309.13608}{Liu (2023)} \cite{Liu:2023ulr} & \( 72.9^{+2.0}_{-2.3} \) & Strong lensing + SnIa \\
14 & \href{https://arxiv.org/abs/2302.13887}{Du (2023)} \cite{Du:2023zsz} & \( 71.5^{+4.4}_{-3.0} \) & Time-delay galaxy lenses + GRBs \\
15 & \href{https://arxiv.org/pdf/2301.09591}{Favale (2023)} \cite{Favale:2023lnp} & \( 71.5 \pm 3.1 \) & Pantheon+ calibrated by Cosmic Chronometers \\
16 & \href{https://arxiv.org/abs/2311.13062}{DESI (2023)} \cite{DESI:2023fij} & \( 85.4^{+29.1}_{-33.9} \) & Dark siren (GW190412 \& DESI) \\
17 & \href{https://arxiv.org/pdf/2305.19914}{Palmese (2023)} \cite{Palmese:2023beh} & \( 75.46^{+5.34}_{-5.39} \) & Standard Siren GW170817 + afterglow \\
18 & \href{https://arxiv.org/abs/2306.09878}{Dominguez (2023)} \cite{Dominguez:2023rxa} & \( 62.4 \pm 4.0 \) & Gamma ray attenuation \\
19 & \href{https://arxiv.org/abs/2306.12468}{Sneppen (2023)} \cite{Sneppen:2023znh} & \( 67.0 \pm 3.6 \) & EPM + Kilonovae \\
20 & \href{https://arxiv.org/pdf/2307.14833}{Liu (2023)} \cite{Liu:2023lbd} & \( 59.1^{+3.6}_{-3.5} \) & Cluster Lensed Quasar \\
21 & \href{https://arxiv.org/pdf/2301.02656}{TDCOSMO (2023)} \cite{TDCOSMO:2023hni} & \( 77.1^{+7.3}_{-7.1} \) & Lensing time delay + stallar kinematics \\
22 & \href{https://arxiv.org/abs/2305.06367}{Kelly (2023)} \cite{Kelly:2023mgv} & \( 66.6 \pm 3.8 \) & Lensing TD SN Refsdal \\
23 & \href{https://arxiv.org/pdf/2307.08285}{Gao (2023)} \cite{Gao:2023izj} & \( 65.5^{+6.4}_{-5.4} \) & Fast Radio Bursts + SnIa \\
24 & \href{https://arxiv.org/pdf/2310.13695}{Alfradique (2023)} \cite{Alfradique:2023giv} & \(  68.84^{+15.5}_{-7.7} \) & 2 new dark sirens (LIGO/Virgo/Delve) \\
25 & \href{https://arxiv.org/pdf/2310.18711}{Colaco (2023)} \cite{Colaco:2023gzy} & \( 67.22\pm 6.07 \) & SZ + X-ray calibrated SnIa  \\
26 & \href{https://arxiv.org/abs/2405.20306}{Pogosian (2024)} \cite{Pogosian:2024ykm} & \( 69.88 \pm 0.93 \) & BAO-DESI data (No Sound Horizon) \\
27 & \href{https://arxiv.org/abs/2405.20306}{Pogosian (2024)} \cite{Pogosian:2024ykm} & \( 67.37 \pm 0.96 \) & BAO-pre DESI data (No Sound Horizon) \\
28 & \href{https://arxiv.org/abs/2402.13476}{Liu (2024)} \cite{Liu:2024onl} & \( 70.0^{+4.7}_{-4.9} \) & Cluster-lensed supernova (SN Refsdal) \\
29 & \href{https://arxiv.org/abs/2404.16092}{Bom (2024)} \cite{Bom:2024afj} & \( 69.9^{+13.3}_{-12.0} \) & Gravitational Waves + Kilonovae \\
30 & \href{https://arxiv.org/abs/2406.02683}{TDCOSMO (2024)} \cite{TDCOSMO:2024rwr} & \( 65^{+23}_{-14} \) & Time-delay cosmography (WGD 2038-4008) \\
31 & \href{https://arxiv.org/abs/2401.10980}{Grillo (2024)} \cite{Grillo:2024rhi} & \( 65.1 \pm 3.5 \) & Lensing TD HFF \\
32 & \href{https://arxiv.org/pdf/2403.18902}{Pascale (2024)} \cite{Pascale:2024qjr} & \( 71.8^{+9.8}_{-7.6} \) & Lensed SnIa (JWST)  \\
33 & \href{https://arxiv.org/pdf/2401.12052}{Li (2024)} \cite{Li:2024elb} & \( 66.3^{+3.8}_{-3.6} \) & Lensed SnIa Refsdal  \\

\bottomrule
\end{tabular}
\caption{Measurements of \( H_0 \) independent of the distance ladder and the sound horizon scale with corresponding first authors and methods.}
\label{tab:independent_h0_measurements}
\end{table*}

1. Reese (2003) \cite{Reese:2003ya}: This analysis presents a measurement of the Hubble constant $H_0$ using the Sunyaev-Zel'dovich effect (SZE) combined with X-ray observations of galaxy clusters. The author reports $H_0 \approx 61 \pm 3 \pm 18$ km s$^{-1}$ Mpc$^{-1}$ for an $\Omega_m = 0.3$, $\Omega_\Lambda = 0.7$ cosmology, where the uncertainties are statistical and systematic at 68\% confidence. The method exploits the different density dependences of the SZE and X-ray emission to determine angular diameter distances to clusters, independent of the cosmic distance ladder. The analysis combines 41 distance determinations to 26 galaxy clusters, using ROSAT X-ray data and modeling clusters as spherical isothermal $\beta$ models. The author critically discusses systematic uncertainties, including cluster structure, gas clumping, and potential SZE contaminants, which dominate the error budget. While the precision is limited by systematics, the technique's ability to measure distances at high redshifts makes it a promising tool for constraining cosmological parameters. The author notes that improved X-ray observations and larger cluster samples could significantly reduce uncertainties in the future.

2. Kuo 2013 \cite{Kuo2013}: This analysis utilized the water megamaser technique to measure the angular diameter distance to NGC 6264, a galaxy located at approximately 140 Mpc. This study is notable for being one of the first to apply the megamaser method to a galaxy well into the Hubble flow, where peculiar velocities have a reduced impact on distance measurements. The authors used Very Long Baseline Interferometry (VLBI) observations to map the spatial and velocity distribution of masers in the circumnuclear disk, combined with single-dish monitoring to measure accelerations. By modeling the three-dimensional structure of the maser disk, they derived a Hubble constant of $H_0 = 68 \pm 9$ km s$^{-1}$ Mpc$^{-1}$. This result is in tension with a later study by Pesce et al. (2020), which found $H_0 = 73.9 \pm 3.0$ km s$^{-1}$ Mpc$^{-1}$ using a larger sample of megamasers. However, the Pesce et al. sample included galaxies at smaller distances, ranging from about 8.5 to 140 Mpc ($z \approx 0.002 - 0.034$), potentially making it more susceptible to biases from peculiar velocities. While the Kuo et al. measurement has larger uncertainties due to the challenges of observing more distant masers, it benefits from reduced systematic errors associated with local flows, making it an important independent check on $H_0$ measurements derived from nearer galaxies.

3. Gao (2015) \cite{Gao:2015tqd}: This analysis presents a measurement of the Hubble constant using water maser observations in the megamaser galaxy NGC 5765b. The authors report $H_0 = 66.0 \pm 6.0$ km s$^{-1}$ Mpc$^{-1}$. The method, part of the Megamaser Cosmology Project. employs very long baseline interferometry (VLBI) to map the spatial distribution of masers and single-dish monitoring to measure their accelerations. These data are combined in a Bayesian framework to fit a warped disk model, simultaneously constraining $H_0$, the black hole mass, and disk geometry. The analysis uses a Markov chain Monte Carlo approach with multiple strands to explore parameter space. Systematic uncertainties, including acceleration measurement methods and unmodeled spiral structure, are carefully considered. The precision achieved (9.1\%) is limited by the acceleration measurements of systemic masers. While not yet competitive with other $H_0$ probes, this geometric method is independent of the cosmic distance ladder and early universe physics. The authors critically discuss the requirements for good distance measurements using this technique and compare their results to previous megamaser analyses. This work demonstrates the potential of megamaser observations to provide independent constraints on $H_0$, which could help resolve the current tension between early and late universe measurements.

4. Millon (2019) and 19. Shajib (2023)\cite{Millon:2019slk,TDCOSMO:2023hni}: Millon et al. (2019) \cite{Millon:2019slk} utilized time-delay cosmography in strong gravitational lensing systems to measure $H_0 = 74.2 \pm 1.6$ km s$^{-1}$ Mpc$^{-1}$. By observing the time delays between multiple images of background quasars lensed by foreground galaxies, and modeling the mass distribution of the lensing galaxies, they provided a measurement independent of the sound horizon scale used in CMB observations. However, subsequent analysis has revealed that this measurement was potentially affected by significant systematic uncertainties.

As highlighted by a subsequent analysis of the same collaboration (TDCOSMO) \citet{TDCOSMO:2023hni}, the analysis by \citet{Millon:2019slk} relied on simple parametric assumptions about the mass profiles of the lensing galaxies, which could potentially bias the $H_0$ measurement or underestimate the errors. The key systematic issue lies in the mass-sheet degeneracy (MSD), which allows for multiple mass distributions to produce the same lensing observables, leading to a degeneracy in the inferred $H_0$ value.

To address these systematics, \citet{TDCOSMO:2023hni} performed a new analysis using spatially resolved stellar kinematics of the lens galaxy RXJ1131$-$1231, obtained from Keck Cosmic Web Imager spectroscopy. This approach allows for a more flexible mass model that effectively breaks the MSD without relying on specific assumptions about the mass profile. By combining this new kinematic data with previously published time delay and lens models derived from Hubble Space Telescope imaging, they were able to robustly estimate $H_0$ while accounting for all uncertainties, including those related to the MSD.

The result of this improved analysis yields $H_0 = 77.1^{+7.3}_{-7.1}$ km s$^{-1}$ Mpc$^{-1}$ for a flat $\Lambda$CDM cosmology. While this new measurement has a larger uncertainty than the previous result from \citet{Millon:2019slk}, it represents a more robust estimate that accounts for the potential systematics introduced by simplified mass profile assumptions. This study demonstrates the importance of incorporating spatially resolved kinematics in time-delay cosmography to mitigate systematic uncertainties and provide a more reliable measurement of the Hubble constant.

5. Pesce (2020) \cite{Pesce:2020xfe}: Pesce et al. (2020) used data from the Megamaser Cosmology Project to derive a new constraint on the Hubble constant of $H_0 = 73.9 \pm 3.0$ km s$^{-1}$ Mpc$^{-1}$. This measurement is independent of distance ladders and the cosmic microwave background. The study combined updated distance measurements for six megamaser-hosting galaxies: UGC 3789, NGC 6264, NGC 6323, NGC 5765b, CGCG 074-064, and NGC 4258.

The authors applied an improved approach for fitting maser data, which incorporated ``error floor'' systematic uncertainties as model parameters. This allowed for marginalization over a previous source of systematic uncertainty and resulted in better distance estimates for four galaxies previously published by the project.

Assuming a fixed velocity uncertainty of 250 km s$^{-1}$ associated with peculiar motions, they derived their best $H_0$ value using only maser-based distance and velocity measurements, without any peculiar velocity corrections. Different approaches for correcting peculiar velocities did not modify $H_0$ by more than $\pm 1\sigma$, with the full range of best-fit Hubble constant values spanning 71.8--76.9 km s$^{-1}$ Mpc$^{-1}$. This study corroborates prior indications that the local value of $H_0$ exceeds the early-Universe value derived from cosmic microwave background measurements. The confidence level for this discrepancy varies from 95--99\% for different treatments of the peculiar velocities.

6. Birrer (2020) \cite{Birrer:2020tax}: In the continuation of leveraging strong lensing systems, Birrer et al. combined the TDCOSMO data set with additional constraints from the SLACS survey, leading to a measurement of \( H_0 = 67.4 \pm 4 \) km/s/Mpc. This work illustrates how adding external data on the lens galaxy's environment and structure, specifically through extended stellar kinematics, can influence the mass profile estimation and thus the derived value of the Hubble constant.

7. Abbott (2021) \cite{LIGOScientific:2021aug}:This analysis presents a comprehensive measurement of the Hubble constant ($H_0$) using gravitational wave (GW) events from GWTC-3. The study employs two main approaches: a hierarchical inference method that jointly estimates cosmological parameters and black hole population properties, and a statistical galaxy catalog method using GLADE+ for redshift information. The hierarchical method, using 42 binary black hole events, yields $H_0 = 68^{+12}_{-8}$ km s$^{-1}$ Mpc$^{-1}$ when combined with GW170817. The galaxy catalog method, incorporating 47 GW events and assuming a fixed black hole population model, results in $H_0 = 68^{+8}_{-6}$ km s$^{-1}$ Mpc$^{-1}$. Both approaches show improvement over previous analyses, with the galaxy catalog method achieving a 42\% improvement compared to GWTC-1 results. However, the authors critically note that the results are dominated by population model assumptions, particularly the mass distribution of black holes. The study highlights the importance of the excess of black holes around 35 $M_\odot$ in constraining $H_0$, but also emphasizes the potential systematic uncertainties introduced by these assumptions. The analysis demonstrates the growing precision of GW cosmology while underscoring the need for careful consideration of population models and systematic effects in future measurements.

8. Wu et al. (2021)\cite{Wu:2021jyk}: This analysis employed a novel method using Fast Radio Bursts (FRBs) to measure the Hubble constant, independent of the cosmic distance ladder. This approach leverages the dispersion measure (DM) of FRBs, which is related to the electron column density along the line of sight and thus serves as a proxy for cosmological distance. The authors used a sample of 18 localized FRBs with known redshifts and carefully modeled the contributions to the DM from the Milky Way, host galaxies, and the intergalactic medium (IGM). By using the DM-redshift relation and accounting for the probability distributions of various DM components, they derived $H_0 = 64.67^{+5.62}_{-4.66}$ km s$^{-1}$ Mpc$^{-1}$. This method is particularly noteworthy as it requires no local distance calibrators and is independent of the cosmic distance ladder, potentially offering a new probe to address the Hubble tension. The authors also demonstrated that with a future sample of 100 localized FRBs, the precision of $H_0$ measurement could reach $\sim$2.6\%, making FRBs a promising tool for cosmology.

9. Philcox (2022) \cite{Philcox:2022sgj}: Philcox et al. introduced a novel approach by using the \( T_{eq} \) standard ruler, independent of the sound horizon scale, deriving \( H_0 = 64.8 \pm 2.4 \) km/s/Mpc. This method utilizes the horizon scale at matter-radiation equality, offering a different sensitivity to early-universe physics compared to traditional \( r_s \)-based measurements, potentially providing insights into discrepancies observed in \( H_0 \) values from early and late universe observations.

10. Zhang (2022) \cite{Zhang:2022caa}: Zhang and colleagues employed cosmic chronometers and HII galaxies, calibrated through a non-parametric Artificial Neural Network, to measure \( H_0 = 65.9 \pm 3.0 \) km/s/Mpc. This method leverages the \( L-\sigma \) relation of HII galaxies, combined with age-dating of cosmic chronometers, to offer an independent and model-agnostic approach to determining the Hubble constant, contributing to resolving the Hubble tension.

11. Bulla (2022) \cite{Bulla:2022ppy}: Utilizing multi-messenger observations from gravitational waves, gamma-ray bursts, and kilonovae associated with neutron star mergers, Bulla et al. reported \( H_0 = 69.6 \pm 5.5 \) km/s/Mpc (weighted mean of the measurements shown in Table 1 of \cite{Bulla:2022ppy} using a typical uncertainty from the Table). This technique harnesses data from different astrophysical events linked by the same originating source, providing a unique way to measure cosmic distances and \( H_0 \) without relying on traditional cosmic ladders. 

12. Moresco (2023) \cite{Moresco:2023zys}: Moresco utilized cosmic chronometers to measure \( H_0 \), obtaining a value of \( 66.7 \pm 5.3 \) km/s/Mpc. This method calculates the Hubble constant by directly determining the age difference between cosmic chronometers at different redshifts, thereby estimating the expansion rate of the Universe independently of traditional distance ladders. This approach provides a unique perspective by focusing on the change in the Universe's age over time to infer \( H_0 \).

13. Liu (2023) \cite{Liu:2023ulr}: This analysis presents a cosmological model-independent measurement of the Hubble constant $H_0$ using strong gravitational lensing time delays combined with Type Ia supernova data. The authors utilize four lensed quasar systems from the H0LiCOW collaboration, which provide time-delay distances and lens angular diameter distances. To avoid assuming a specific cosmological model, they reconstruct the distance-redshift relation from the Pantheon SN Ia sample using Gaussian process regression. By combining these unanchored distances with the lensing data, they simultaneously constrain $H_0$ and the post-Newtonian parameter $\gamma_{PPN}$. Their main result is $H_0 = 72.9^{+2.0}_{-2.3}$ km s$^{-1}$ Mpc$^{-1}$. The method is robust as it does not rely on a parametric cosmological model, potentially reducing biases. However, it still assumes spatial flatness and depends on the chosen mean function for the Gaussian process. The precision is comparable to model-dependent analyses, demonstrating the power of combining lensing time delays with standardizable candles to measure $H_0$ independently of the cosmic distance ladder or early universe physics.

14. Du (2023) \cite{Du:2023zsz}: This analysis presents a model-independent determination of the Hubble constant $H_0$ and spatial curvature $\Omega_{K,0}$ using time-delay galaxy lenses (TDGLs) combined with gamma-ray burst (GRB) distances. The authors employ the distance sum rule, assuming only the validity of the FLRW metric and geometrical optics. They use 7 TDGLs from H0LiCOW and STRIDES collaborations, along with 193 GRBs calibrated via the Amati relation. The method parameterizes distances using a fourth-order cosmographic expansion in y-redshift, including curvature. Combining TDGLs and GRBs yields $H_0 = 71.5^{+4.4}_{-3.0}$ km s$^{-1}$ Mpc$^{-1}$ and $\Omega_{K,0} = -0.07^{+0.13}_{-0.06}$. Assuming a flat universe improves the constraint to $H_0 = 70.9^{+4.2}_{-2.9}$ km s$^{-1}$ Mpc$^{-1}$. The analysis is performed in a Bayesian framework using dynamic nested sampling. While the precision is not yet competitive with other probes, this method provides a novel, independent constraint on both $H_0$ and $\Omega_{K,0}$. The authors critically discuss potential systematic effects, including GRB standardization and the limited number of low-redshift GRBs. They also explore future prospects using simulated GRB samples, demonstrating potential for significant improvement in precision.

15. Favale (2023) \cite{Favale:2023lnp}: This analysis employs cosmic chronometers (CCH), Type Ia supernovae (SNIa) from the Pantheon+ compilation, and baryon acoustic oscillation (BAO) data to perform a model-independent calibration of the cosmic distance ladder and measurement of the Hubble constant $H_0$. The authors use Gaussian Process regression to reconstruct $H(z)$ from CCH data and $m(z)$ from SNIa data. They then combine these reconstructions with BAO data to jointly constrain the absolute magnitude of SNIa $M$, the sound horizon scale $r_d$, and the curvature parameter $\Omega_k$. Using CCH+SNIa+BAO data (excluding SNIa in SH0ES host galaxies), they obtain $H_0 = 71.5 \pm 3.1$ km s$^{-1}$ Mpc$^{-1}$. The method is largely independent of early universe physics and the first rungs of the cosmic distance ladder, providing a valuable cross-check on more model-dependent approaches. However, the precision is still limited compared to other methods, and the result falls between the values preferred by Planck and SH0ES, unable to definitively resolve the Hubble tension. The authors also explore the impact of potentially overestimated CCH uncertainties and the inclusion of SH0ES host galaxy SNIa, demonstrating the robustness of their approach while highlighting areas for future improvement as more precise data become available.

16. DESI (2023) \cite{DESI:2023fij}: This analysis presents a measurement of the Hubble constant $H_0$ using the gravitational wave event GW190412 as a dark standard siren, combined with galaxy data from the Dark Energy Spectroscopic Instrument (DESI). The authors employ the statistical standard siren method, marginalizing over potential host galaxies from DESI's Bright Galaxy Sample. They find $H_0 = 85.4^{+29.1}_{-33.9}$ km s$^{-1}$ Mpc$^{-1}$ for GW190412 alone, and $H_0 = 77.96^{+18.4}_{-9.48}$ km s$^{-1}$ Mpc$^{-1}$ when combined with the bright siren GW170817. The analysis uses a flat $\Lambda$CDM cosmology with $\Omega_m = 0.3$ and a uniform prior on $H_0$ between 20 and 140 km s$^{-1}$ Mpc$^{-1}$. The method critically relies on the completeness and accuracy of the DESI spectroscopic sample, which offers significant improvements over previous photometric catalogs. The authors explore the impact of various large-scale structure weights and different data releases. While the precision is not yet competitive with other probes, this work demonstrates the potential of combining well-localized gravitational wave events with deep spectroscopic surveys. The multi-peaked posterior reflects redshift overdensities in the GW localization volume, highlighting the importance of accurate redshift information in dark siren analyses.

17. Palmese (2023) \cite{Palmese:2023beh}: This analysis presents an updated measurement of the Hubble constant $H_0$ using GW170817 and the latest observations of its electromagnetic counterpart afterglow. The authors employ a Bayesian framework that combines gravitational wave data, afterglow observations spanning ~3.5 years across multiple wavelengths, and a careful treatment of the host galaxy's peculiar velocity. They use the JetFit package to constrain the viewing angle from afterglow data, which helps break the distance-inclination degeneracy in the GW data. The analysis yields $H_0 = 75.46^{+5.34}_{-5.39}$ km s$^{-1}$ Mpc$^{-1}$ (68\% credible interval), a ~7\% precision measurement that improves significantly upon the original 14\% precision from the first standard siren measurement. The authors critically examine various assumptions in their modeling, particularly the impact of the bulk Lorentz factor prior on the viewing angle and $H_0$ estimates. While the result is consistent within 1$\sigma$ of the Cepheid-anchored supernova measurements and within 1.5$\sigma$ of CMB measurements, the precision is not yet sufficient to resolve the Hubble tension. The study demonstrates the potential of long-term, multi-wavelength afterglow observations in improving standard siren measurements, but also highlights the need for careful consideration of modeling assumptions in future high-precision cosmological studies using this method.

18. Dominguez (2023) \cite{Dominguez:2023rxa}: This analysis presents a measurement of the Hubble constant $H_0$ using $\gamma$-ray attenuation data combined with an extragalactic background light (EBL) model derived from galaxy observations. The authors obtain $H_0 = 65.1^{+6.0}_{-4.9}$ km s$^{-1}$ Mpc$^{-1}$ when simultaneously fitting $H_0$ and $\Omega_m$, and $H_0 = 62.4^{+4.1}_{-3.9}$ km s$^{-1}$ Mpc$^{-1}$ when fixing $\Omega_m = 0.32$. The method uses optical depth measurements from Fermi-LAT and Imaging Atmospheric Cherenkov Telescopes, combined with an EBL model based on deep multiwavelength galaxy data from CANDELS. The analysis accounts for cosmological dependence in both the EBL model and the line-of-sight integral for optical depth. Uncertainties in the EBL model are propagated through Monte Carlo realizations. The results are consistent with Planck CMB measurements but in tension with local distance ladder measurements. The authors critically discuss potential systematic effects, including discrepancies between EBL models and $\gamma$-ray data at high redshifts, and explore the impact of restricting the analysis to low-redshift data. While the method provides an independent probe of $H_0$, current uncertainties are too large to definitively resolve the Hubble tension.

19. Sneppen (2023) \cite{Sneppen:2023znh}: This analysis presents a measurement of the Hubble constant $H_0$ using the expanding photosphere method (EPM) applied to the kilonova AT2017gfo. The authors analyze early VLT/X-shooter spectra, modeling the continuum as a blackbody and the Sr II lines as P Cygni profiles. They find $H_0 = 67.0 \pm 3.6$ km s$^{-1}$ Mpc$^{-1}$ by combining constraints from the first two epochs. The method exploits the well-defined explosion time from gravitational wave detection, the apparent sphericity of the ejecta, and the negligible electron scattering opacity in kilonova atmospheres. The dominant uncertainty comes from the host galaxy peculiar velocity, while systematic uncertainties related to dust extinction, flux calibration, and ejecta geometry contribute at the ~1-2\% level each. The authors critically examine potential systematics, including line blending, reverberation effects, and relativistic corrections. While demonstrating the potential of the method, they caution that their simple framework does not account for full 3D radiative transfer effects, and the result should be considered preliminary. They forecast that 5-10 similar events could constrain $H_0$ to ~1\% precision, potentially helping to resolve the Hubble tension.

20. Liu (2023) \cite{Liu:2023lbd}: This analysis explores the measurement of the Hubble constant $H_0$ using time-delay cosmography with the cluster-lensed quasar system SDSS J1004+4112. The authors employ 16 different lens mass models to investigate the dependence of $H_0$ on model assumptions. By combining posteriors from all models with equal weighting, they obtain $H_0 = 67.5^{+14.5}_{-8.9}$ km s$^{-1}$ Mpc$^{-1}$. The study critically examines the impact of various modeling choices, including dark matter halo profiles, treatment of the brightest cluster galaxy, and multipole perturbations. They find that $H_0$ values decrease as model complexity increases, and that different halo profiles yield significantly different results. The authors also explore weighting schemes and additional constraints to improve the measurement. By selecting two models that best reproduce observed shapes of lensed galaxies, they obtain a tighter constraint of $H_0 = 59.1^{+3.6}_{-3.5}$ km s$^{-1}$ Mpc$^{-1}$. However, they caution that their analysis does not fully explore all possible mass model uncertainties. The study highlights the importance of including extended arc information in cluster lens modeling for obtaining accurate $H_0$ constraints, while also emphasizing the challenges in rigorously incorporating such information due to small-scale matter distributions affecting image positions.

21. TDCOSMO (2023) \cite{TDCOSMO:2023hni}: (see 3. above)

22. Kelly (2023) \cite{Kelly:2023mgv}: Kelly et al. employed time-delay strong lensing, specifically focusing on SN Refsdal, to measure \( H_0 = 66.6 \pm 3.8 \) km/s/Mpc. This method utilizes the time delays between multiple observed images of a single supernova event caused by gravitational lensing. The precision of the measurement benefits from the unique opportunity provided by observing multiple light curves of the same supernova, allowing for a detailed modeling of the lensing mass distribution and the overall geometry of the system.

23. Gao (2023) \cite{Gao:2023izj}: This analysis employs a cosmographic approach to measure the Hubble constant $H_0$ using Fast Radio Bursts (FRBs) and Type Ia supernovae (SnIa). The authors derive a Taylor expansion for the mean intergalactic medium dispersion measure $\langle\text{DM}_\text{IGM}(z)\rangle$ of FRBs and combine it with uncalibrated SnIa data from the Pantheon sample to constrain cosmographic parameters. Using 18 localized FRBs and assuming a constant baryon fraction in the IGM, they obtain $H_0 = 65.5^{+6.4}_{-5.4}$ km s$^{-1}$ Mpc$^{-1}$ (68\% C.L.). The method is largely independent of cosmological models but relies on assumptions about the IGM baryon fraction and FRB host galaxy properties. The authors critically examine the impact of these assumptions and parameter degeneracies on their results. They find that allowing for redshift evolution in the IGM baryon fraction shifts the $H_0$ estimate to $69.0^{+6.7}_{-5.7}$ km s$^{-1}$ Mpc$^{-1}$. While the precision is lower than some other recent FRB-based measurements, this approach provides a useful cross-check that is less dependent on early-universe physics. The authors also present forecasts suggesting that $\sim$100 localized FRBs could improve the $H_0$ precision to $\sim$4.6\%.

24. Alfradique (2023) \cite{Alfradique:2023giv}: This analysis presents a measurement of the Hubble constant $H_0$ using the dark siren method with two new gravitational wave (GW) events, GW190924-021846 and GW200202-154313, combined with previous results. The authors employ the DECam Local Volume Exploration Survey (DELVE) galaxy catalog to obtain photometric redshifts of potential host galaxies using a Mixture Density Network approach. They implement a Bayesian framework that marginalizes over possible host galaxies and incorporates photometric redshift uncertainties. The combination of these two new events with the bright siren GW170817 yields $H_0 = 68.84^{+15.51}_{-7.74}$ km s$^{-1}$ Mpc$^{-1}$, improving the precision by 7\% compared to GW170817 alone. When combined with 8 previously analyzed dark sirens, they obtain $H_0 = 76.00^{+17.64}_{-13.45}$ km s$^{-1}$ Mpc$^{-1}$. The analysis critically examines the impact of photometric redshift bias and the use of full redshift probability distributions. While the precision is lower than some recent studies using larger GW samples, this approach is less sensitive to assumptions about the black hole mass distribution and demonstrates the potential of well-localized dark sirens for cosmological measurements. The authors note that future analyses will need to address additional systematic effects as the precision improves. 

25. Colaco (2023) \cite{Colaco:2023gzy}: This analysis presents a measurement of the Hubble constant $H_0$ using a combination of galaxy cluster angular diameter distances and type Ia supernovae (SnIa) luminosity distances. The method employs a model-independent approach based on the cosmic distance duality relation, requiring only geometrical distances. The authors use 25 galaxy cluster measurements from De Filippis et al. (2005), modeled with an elliptical $\beta$-model, and the Pantheon SnIa sample. A Gaussian Process regression is applied to reconstruct the SnIa distance-redshift relation at galaxy cluster redshifts. The analysis yields $H_0 = 67.2 \pm 6.1$ km s$^{-1}$ Mpc$^{-1}$ (68\% confidence level), with an uncertainty of about 9\%. This result is consistent with both Planck CMB and local distance ladder measurements, though it does not resolve the Hubble tension. The authors critically discuss the impact of cluster modeling assumptions and combine statistical and systematic errors in quadrature. While the precision is limited by current galaxy cluster measurements, this approach provides a promising avenue for future constraints as cluster data improve, offering an independent probe of $H_0$ that avoids potential biases from local distance calibration or specific cosmological model assumptions.

26-27. Pogosian (2024) \cite{Pogosian:2024ykm}: This analysis presents a cosmological-model-independent measurement of the Hubble constant $H_0$ using the DESI Year 1 BAO data combined with the CMB acoustic scale $\theta_*$ and a Planck $\Lambda$CDM prior on $\Omega_m h^2$. The authors treat the sound horizon at baryon decoupling $r_d$ as a free parameter, avoiding assumptions about recombination physics. Using CosmoMC, they find $H_0 = 69.88 \pm 0.93$ km s$^{-1}$ Mpc$^{-1}$, which is $\sim 2\sigma$ higher than the Planck $\Lambda$CDM value. For comparison, they perform the same analysis with pre-DESI BAO data, obtaining $H_0 = 67.37 \pm 0.96$ km s$^{-1}$ Mpc$^{-1}$. The difference stems from the larger $r_d h$ value measured by DESI. This method provides a valuable consistency test of the cosmological model at recombination, with no direct dependence on the sound horizon scale. It should be stressed however that the assumed prior on $\Omega_m h^2$ is model dependent assumes \lcdm and would be violated in the context of EDE models which predict a higher value of $\Omega_m h^2$\cite{Poulin:2023lkg}. The authors critically discuss the differences between DESI and pre-DESI results, noting that future DESI data covering larger volumes will help resolve discrepancies. While this approach does not fully resolve the Hubble tension, it offers important insights into potential modifications needed in the standard cosmological model around the epoch of recombination.

28. Liu (2024) \cite{Liu:2024onl}: This analysis presents a detailed study of the Hubble constant ($H_0$) measurement using the cluster-lensed supernova SN Refsdal in MACS J1149.5+2223. The authors employ 23 different lens mass models to thoroughly explore the dependence of $H_0$ on mass model assumptions. Using the software glafic, they construct parametric models with varying dark matter halo profiles, components, and multipole perturbations. By combining measurements from all models with equal weighting, they obtain $H_0 = 70.0^{+4.7}_{-4.9}$ km s$^{-1}$ Mpc$^{-1}$. A weighting scheme based on positional errors yields a slightly tighter constraint of $H_0 = 70.3 \pm 4.4$ km s$^{-1}$ Mpc$^{-1}$. The analysis reveals that the $H_0$ measurement is relatively robust against different mass model assumptions, unlike previous studies on cluster-lensed quasars. The authors critically examine correlations between $H_0$ and various lensing properties, finding that steeper radial density profiles tend to yield larger $H_0$ values. They also demonstrate a clear correlation between $H_0$ and magnification factors, highlighting the potential of gravitationally lensed Type Ia supernovae for accurate $H_0$ measurements. While the precision is comparable to previous analyses of SN Refsdal, this study provides a comprehensive exploration of mass model dependencies and insights for future improvements in $H_0$ constraints from cluster lensing.

29. Bom (2024) \cite{Bom:2024afj}: This analysis presents a new measurement of the Hubble constant $H_0$ using the dark standard siren method with 5 well-localized gravitational wave events from the LIGO/Virgo/KAGRA O4a observing run, combined with 10 events from previous runs. The authors employ the galaxy catalog method, using photometric redshifts derived from deep learning techniques applied to DESI Legacy Survey and DELVE data. They find $H_0 = 69.9^{+13.3}_{-12.0}$ km s$^{-1}$ Mpc$^{-1}$ (68\% HDI) from 15 dark sirens alone, and $H_0 = 68.0^{+4.3}_{-3.8}$ km s$^{-1}$ Mpc$^{-1}$ when combined with the bright siren GW170817 and its electromagnetic counterpart. The analysis critically examines the impact of photometric redshift quality, catalog completeness, and selection effects. The authors use a Bayesian framework to marginalize over potential host galaxies and incorporate a detailed treatment of the gravitational wave selection function. While the precision is comparable to previous dark siren analyses, this work demonstrates the potential of combining multiple well-localized events with high-quality photometric redshifts. The authors discuss potential systematics and limitations, particularly related to BBH population assumptions and catalog depth. This approach provides an independent probe of $H_0$ that could help arbitrate the Hubble tension as more events are detected and analyzed.

30. TDCOSMO (2024) \cite{TDCOSMO:2024rwr}: This analysis presents a measurement of the Hubble constant $H_0$ using time-delay cosmography applied to the quadruply imaged lensed quasar WGD 2038 4008. The authors combine new time-delay measurements from optical monitoring with existing lens models to constrain the time-delay distance $D_{\Delta t}$ and $H_0$. They employ two independent lens modeling codes (GLEE and Lenstronomy) and consider both power-law and composite (stars+dark matter) mass profiles. The analysis incorporates the full time-delay covariance matrix and external convergence estimates. Assuming a flat $\Lambda$CDM cosmology with $\Omega_m = 0.3$, they find $H_0 = 65^{+23}_{-14}$ km s$^{-1}$ Mpc$^{-1}$. The large uncertainty is primarily due to the weak time-delay constraints, as the quasar exhibited low variability during the monitoring period. The analysis was performed blindly with respect to cosmological parameters to prevent experimenter bias. This is the first lens in the TDCOSMO sample to incorporate multiple lens models in the final inference and use the full time-delay covariance matrix. While the precision is insufficient to address the Hubble tension, this system will contribute to future joint analyses of multiple lenses to improve overall constraints on $H_0$.

\begin{figure*}[ht]
\centering
\includegraphics[width=\textwidth]{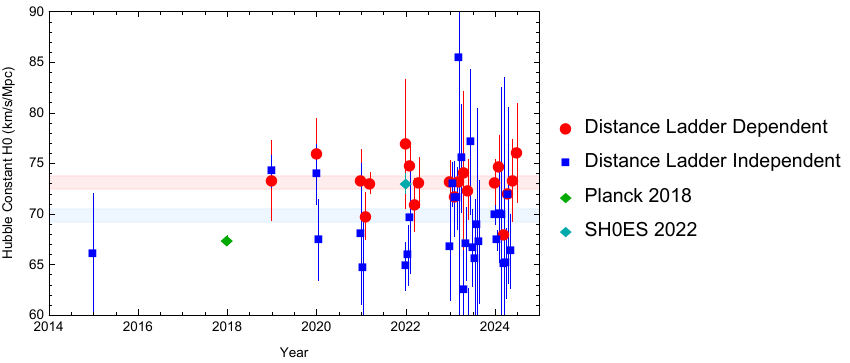}
\caption{The measurements of $H_0$ shown in Tables \ref{tab:distance_ladder_h0_measurements} (distance ladder dependent-red points) and \ref{tab:independent_h0_measurements}
 (one step distance ladder and sound horizon independent-blue points). The most precise distance ladder dependent measurement (SH0ES 2022\cite{Riess:2021jrx}) and the most precise sound horizon dependent Planck 18\cite{Planck:2018vyg} measurements are also shown as green points. The light red and light blue bands denote the combined constraints for distance ladder dependent and one step measurements respectively.} \label{fig:hubble_constant}
\end{figure*}

31. Grillo (2024) \cite{Grillo:2024rhi}: This analysis presents a measurement of the Hubble constant $H_0$ using strong gravitational lensing in the galaxy cluster MACS J1149.5+2223, including time delays of the multiply-imaged supernova Refsdal. The authors employ a parametric lens model constrained by 89 multiple images from 28 background sources and 4 measured time delays of SN Refsdal. In an open wCDM cosmology, they find $H_0 = 65.1^{+3.5}_{-3.4}$ km s$^{-1}$ Mpc$^{-1}$. The analysis is performed using the GLEE software, considering four cosmological models (flat and open $\Lambda$CDM and wCDM) and three lens model variants to assess systematic uncertainties. Remarkably, the $H_0$ measurement is robust across different cosmological models and lens parameterizations, with systematics estimated to be smaller than statistical errors. The precision achieved (~6\%) is competitive with other probes, and the method uniquely constrains multiple cosmological parameters simultaneously without external priors. The authors critically compare their results to other cosmological probes, demonstrating the complementarity and stability of their method. While based on a single lens system, this work showcases the potential of time-delay cosmography with galaxy clusters as a powerful and independent probe of cosmology.

32. Pascale (2024) \cite{Pascale:2024qjr}: This analysis presents the first precision measurement of the Hubble constant $H_0$ using a multiply-imaged Type Ia supernova, SN H0pe, discovered by JWST in the galaxy cluster PLCK G165.7+67.0. The authors employ a Bayesian approach combining photometric and spectroscopic time delay measurements with seven independent lens models. The standard candle nature of SN Ia allows for absolute magnification constraints, which are used to break degeneracies between lens models. Following strict blinding protocols, they obtain $H_0 = 71.8^{+9.8}_{-7.6}$ km s$^{-1}$ Mpc$^{-1}$. The analysis critically examines the impact of different lens modeling approaches, the correlation between predicted time delays and magnifications, and the effect of including weak lensing data. The authors discuss potential systematics, including the interpolation method for the unlensed SN Ia apparent magnitude and the thin-lens approximation. While the precision is limited by both lens modeling and time delay measurement uncertainties, this work demonstrates the power of cluster-lensed SnIa for time-delay cosmography. The result is consistent with local universe measurements but in tension with early universe predictions at the $\sim$1.5$\sigma$ level, providing new evidence in the context of the Hubble tension.

33. Li (2024) \cite{Li:2024elb}: This analysis presents a cosmological-model-independent measurement of the Hubble constant $H_0$ using the gravitationally lensed supernova Refsdal in combination with the Pantheon+ supernova sample. The authors employ Gaussian process regression to reconstruct the expansion history from supernovae Ia data, which is then anchored by the time-delay distance of SN Refsdal. Using eight lens models for the cluster MACS J1149, they infer $H_0 = 64.2^{+4.4}_{-4.3}$ km s$^{-1}$ Mpc$^{-1}$, while using only the two models most consistent with observations yields $H_0 = 66.3^{+3.8}_{-3.6}$ km s$^{-1}$ Mpc$^{-1}$. These results are consistent with previous model-dependent analyses but avoid potential biases from assuming a specific cosmological model. The precision is comparable to model-dependent methods, demonstrating the power of this approach. The results are in tension with distance ladder measurements at the 1.8-2$\sigma$ level but agree well with Planck CMB data. The authors critically compare their findings to other time-delay cosmography results and discuss the impact of lens modeling uncertainties. While assuming spatial flatness, this method provides a valuable cross-check on $H_0$ measurements that is less sensitive to cosmological model assumptions.

\begin{figure*}[ht]
\centering
\includegraphics[width=\textwidth]{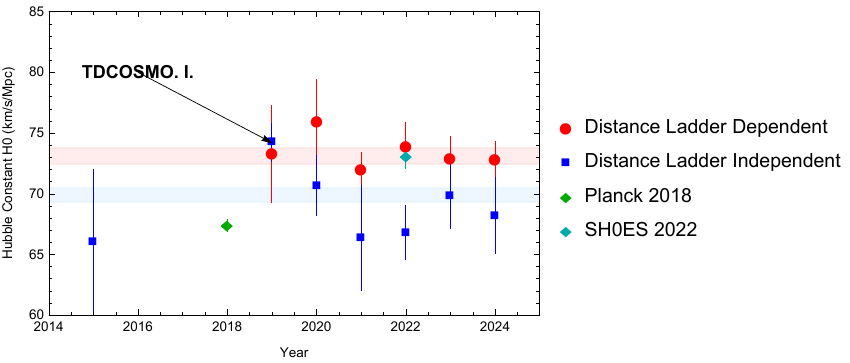}
\caption{Same as Fig. \ref{fig:hubble_constant} but the measurements are binned by year. Notice that for most years the one step $H_0$ are consistent with the Planck 18 measurement of $H_0$. The only one step measurement that is in significant tension with Planck 18\cite{Planck:2018vyg} is the TDCOSMO. I. strong lensing time delay measurement\cite{Millon:2019slk} which has been shown to suffer from a systematic issue (related to 
 the mass-sheet degeneracy (MSD)) by  later TDCOSMO analyses\cite{TDCOSMO:2023hni,Birrer:2020tax}. 
However, more recent TDCOSMO measurements are consistent with Planck 18 and other sound horizon based measurements (eg \cite{Birrer:2020tax,TDCOSMO:2024rwr}).}
\label{fig:hubble_constant_binned}
\end{figure*}

\begin{figure*}
    \includegraphics[width=\textwidth]{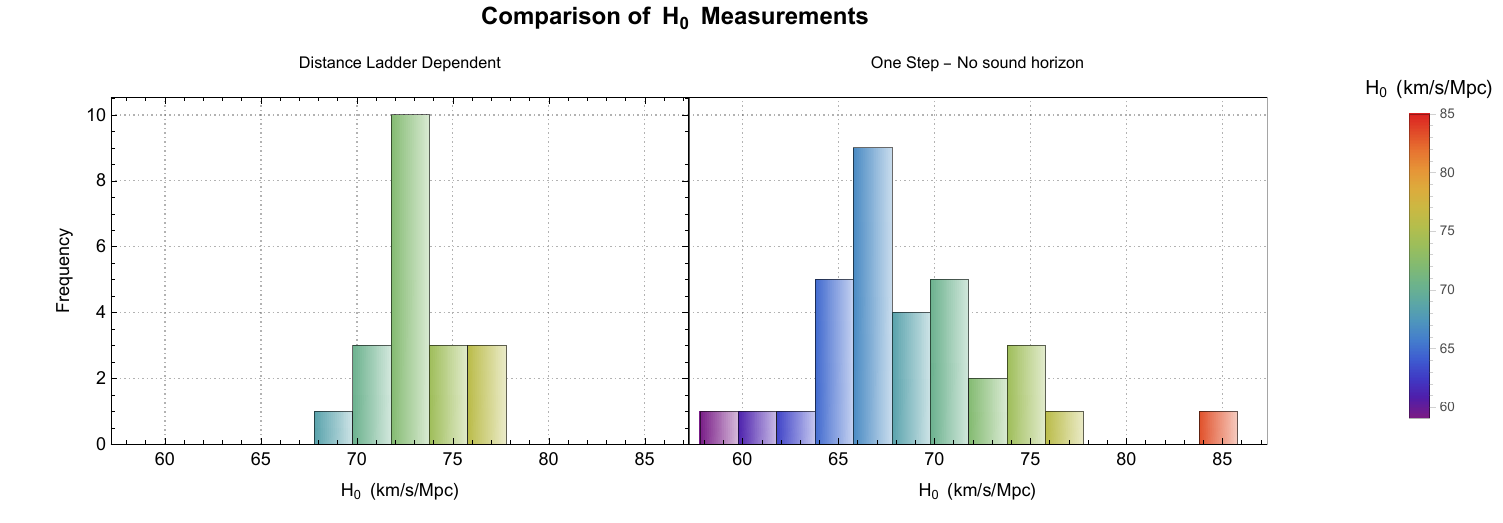}
    \caption{Histogram comparison of $H_0$ measurements. Left: Distribution of Distance Ladder Dependent measurements. Right: Distribution of One Step - No sound horizon measurements. The color gradient represents the range of $H_0$ values from approximately 60 to 85 km/s/Mpc. This comparison highlights the frequency distribution of measurements for each method, allowing for a direct visual comparison of their respective ranges and central tendencies.}
    \label{fig:histogram_comparison}
\end{figure*}

\begin{figure}
    \includegraphics[width=\columnwidth]{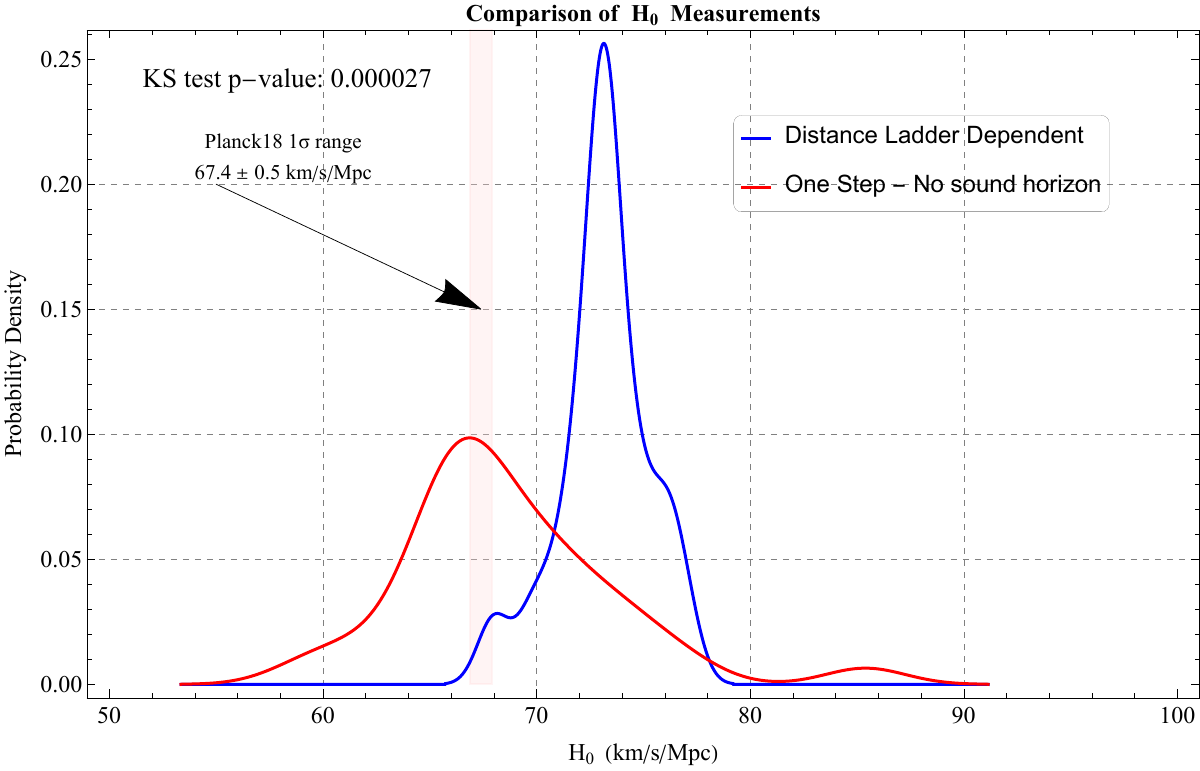}
    \caption{Comparison of $H_0$ measurements using smooth histogram representations. The blue curve represents the Distance Ladder Dependent measurements, while the red curve shows the One Step - No sound horizon measurements. The vertical light red band indicates the Planck18 $1\sigma$ range of $67.4 \pm 0.5$ km/s/Mpc. A Kolmogorov-Smirnov test between the two distributions yields a p-value of 0.000102, indicating a statistically significant difference between the two measurement methods.}
    \label{fig:smooth_histogram}
\end{figure}

\subsection{Results: One step methods vs distance ladder methods}

In this section, we present a comprehensive statistical analysis of the Hubble constant ($H_0$) measurements, categorized into two distinct groups: distance ladder-dependent measurements and one-step measurements that are independent of both the distance ladder and the sound horizon scale. Our analysis reveals significant differences between these two approaches, with important implications for the Hubble tension.

Figures \ref{fig:hubble_constant} and \ref{fig:hubble_constant_binned} provide a visual representation of the $H_0$ measurements from both categories. In Figure \ref{fig:hubble_constant}, we observe a clear separation between the distance ladder-dependent measurements (red points) and the one-step measurements (blue points). The distance ladder measurements consistently yield higher $H_0$ values, clustering around 73 km/s/Mpc, while the one-step measurements tend to be lower, grouping closer to the Planck 18 value of $67.4 \pm 0.5$ km/s/Mpc. 

Figure \ref{fig:hubble_constant_binned} presents the same data binned by year, offering insights into the temporal evolution of $H_0$ measurements. This visualization highlights a consistent trend over recent years: one-step measurements (blue points) consistently report lower $H_0$ values compared to distance ladder measurements (red points). Notably, most one-step measurements align well with the Planck 18 result, with the exception of the 2019 TDCOSMO.I strong lensing time delay measurement. This outlier has since been attributed to systematic issues related to the mass-sheet degeneracy (MSD), as demonstrated by subsequent TDCOSMO analyses.

The histogram comparison in Figure \ref{fig:histogram_comparison} further illustrates the distinct distributions of $H_0$ measurements between the two methods. The distance ladder-dependent measurements (left panel) show a distribution centered at higher $H_0$ values, while the one-step measurements (right panel) exhibit a distribution shifted towards lower values, more consistent with Planck 18 results.

Figure \ref{fig:smooth_histogram} presents a smooth histogram representation of the two measurement categories, providing a clear visualization of their distinct probability distributions. The blue curve (one-step measurements) shows a peak that aligns closely with the Planck 18 $1\sigma$ range (indicated by the vertical light red band), while the red curve (distance ladder-dependent measurements) peaks at a notably higher $H_0$ value.

To combine the multiple measurements of $H_0$ from both distance ladder dependent and independent methods, we employed a weighted average approach. This method assumes that all measurements are independent of each other. The procedure is as follows:

\begin{enumerate}
    \item \textbf{Weighted Mean Calculation:} 
    Each $H_0$ measurement is weighted by the inverse square of its uncertainty, giving more weight to measurements with smaller uncertainties. The weighted mean $H_0$ is calculated as:
    
    \begin{equation}
        \overline{H_0} = \frac{\sum_i H_{0,i} / \sigma_i^2}{\sum_i 1 / \sigma_i^2}
    \end{equation}
    
    where $H_{0,i}$ is the $i$-th measurement and $\sigma_i$ is its uncertainty.

    \item \textbf{Uncertainty Calculation:} 
    The uncertainty of the weighted mean is estimated as the inverse square root of the sum of weights:
    
    \begin{equation}
        \sigma_{\overline{H_0}} = \frac{1}{\sqrt{\sum_i 1 / \sigma_i^2}}
    \end{equation}
\end{enumerate}
This method assumes that all measurements are independent of each other, allowing us to treat each measurement as providing separate information. However, it's important to note that in reality, some measurements might have correlations due to shared data or methodologies. The independence assumption, while commonly used for simplicity, may lead to an underestimation of the final uncertainty if significant correlations exist between measurements.  

Our statistical analysis reveals significant differences between the two groups of $H_0$ measurements:

\begin{enumerate}
    \item \textbf{Weighted Means:} 
    \begin{itemize}
        \item Distance Ladder Dependent: $H_0 = 72.8 \pm 0.5$ km/s/Mpc
        \item One-step (Distance Ladder Independent): $H_0 = 69.0 \pm 0.5$ km/s/Mpc. This value decreases further to $H_0 = 68.3 \pm 0.5$ km/s/Mpc if two outliers are removed from the sample as discussed below.
    \end{itemize}
    These weighted means differ by approximately 3.75 km/s/Mpc, which is statistically significant given their respective uncertainties.

    \item \textbf{Distribution Comparison:} A Kolmogorov-Smirnov (KS) test comparing the two groups of measurements yields a p-value of 0.000102. This extremely low p-value indicates that the probability of these two distributions being drawn from the same underlying population is negligible. In other words, there is strong statistical evidence that the distance ladder-dependent and one-step measurements represent fundamentally different distributions of $H_0$ values.
\end{enumerate}

The consistency of one-step measurements with sound horizon-based measurements like Planck 18 ($H_0 = 67.4 \pm 0.5$ km/s/Mpc) and standard BAO measurements is particularly noteworthy. This alignment suggests that the one-step methods, which are independent of both the distance ladder and sound horizon scale, corroborate the lower $H_0$ values inferred from early universe observations.

\subsubsection{$\chi^2$ analysis and robustness of results}

To further assess the consistency and reliability of our findings, we conducted a comprehensive $\chi^2$ analysis on various subsets of the distance ladder independent dataset. This analysis serves to evaluate the internal consistency of different measurement methods and identify potential sources of tension within the data.

We constructed and minimized the $\chi^2$ statistic for each subset, defined as:

\begin{equation}
\chi^2 = \sum_i \frac{(H_{0,i} - H_{0,\text{best fit}})^2}{\sigma_i^2}
\end{equation}
where $H_{0,i}$ are the individual measurements, $\sigma_i$ their uncertainties, and $H_{0,\text{best fit}}$ the best-fit value that minimizes $\chi^2$.

Table \ref{tab:chi_square_analysis} presents the results of this analysis for various subsets of the data:

\begin{table*}
\centering
\caption{Analysis of Various Subsets of the One Step - Sound Horizon and Distance Ladder Independent Sample. The data numbers correspond to Table \ref{tab:independent_h0_measurements}. Notice the dramatic improvement of the self consistency of the full sample (decrease of the reduced $\chi^2_\text{red}\equiv \chi^2/dof$ ) if the two outliers (TDCOSMO.I and MCP-SH0ES) are removed. These outliers have been removed from most of the subsamples after row 2. If they were present the value of $\chi^2_{red}$ would have been significantly larger (see eg row 5 where TDCOSMO.I has been included).}
\label{tab:chi_square_analysis}
\begin{tabular}{lccccc}
\hline\hline
Dataset & $H_0$ (km s$^{-1}$ Mpc$^{-1}$) & $\chi^2_\text{min}$ & $\chi^2_\text{red}$ & DoF & Data points \\
\hline
Full Sample & $69.0 \pm 0.5$ & 43.69 & 1.37 & 32 & All \\
Full Sample (excl. 4,5) & $68.3 \pm 0.5$ & 28.51 & 0.95 & 30 & All except outliers 4,5 \\
Gravitational Waves (GW) & $71.4 \pm 3.1$ & 1.18 & 0.24 & 5 & 7,11,16,17,24,29 \\
Full Sample (excl. GW) & $68.2 \pm 0.5$ & 26.31 & 1.10 & 24 & All except 4,5,7,11,16,17,24,29 \\
TD Lensing All & $71.4 \pm 1.0$ & 12.00 & 1.33 & 9 & 4,6,13,14,21,22,28,31-33 \\
Full Sample (excl. TD Lensing) & $68.2 \pm 0.6$ & 23.46 & 1.07 & 22 & All except 4-6,13,14,21,22,28,31-33 \\
TD Lensing (no TDCOSMO I) & $69.7 \pm 1.2$ & 7.14 & 0.89 & 8 & 6,13,14,21,22,28,31-33 \\
Cosmic Chronometers & $68.3 \pm 2.0$ & 1.80 & 0.90 & 2 & 10,12,15 \\
Full Sample (excl. CC) & $68.3 \pm 0.5$ & 26.71 & 0.99 & 27 & All except 4,5,10,12,15 \\
Megamasers & $72.0 \pm 2.6$ & 1.60 & 0.80 & 2 & 2,3,5 \\
Full Sample (excl. Megamasers) & $68.3 \pm 0.5$ & 28.36 & 1.01 & 28 & All except 2-5 \\
Megamasers (excl. MCP-SH0ES) & $67.0 \pm 5.0$ & 0.03 & 0.03 & 1 & 2,3 \\
Early Time (No Sound Horizon) & $68.4 \pm 0.6$ & 5.93 & 2.97 & 2 & 9,26,27 \\
Full Sample (excl. Early Time) & $68.1 \pm 0.9$ & 22.50 & 0.83 & 27 & All except 4,5,9,26,27 \\
\hline\hline
\end{tabular}
\end{table*}

Our analysis reveals that most subsamples are self-consistent and in agreement with each other, as evidenced by their reduced $\chi^2$ values close to 1. However, we observe some tension in the full dataset (reduced $\chi^2 = 1.43$) and in two subsamples: Megamasers and Lensing All.

The tension in the full dataset and the in the above two subsamples can be attributed to two notable outliers: data points 3 and 4 of Table \ref{tab:independent_h0_measurements}, associated with TDCOSMO I lensing\cite{Millon:2019slk} and the Megamasers MCP+SH0ES \cite{Pesce:2020xfe} points respectively. The TDCOSMO I measurement (point 3) has been shown to suffer from systematic issues related to the mass-sheet degeneracy (MSD), as demonstrated by subsequent TDCOSMO analyses\cite{TDCOSMO:2023hni,Birrer:2020tax}. When we exclude these two outliers (Full Data excl. 3,4), the reduced $\chi^2$ ($\chi^2$ per dof) improves significantly to 1.00585, indicating excellent consistency among the remaining data points. In addition, the consistency of the remaining one step-sound horizon free measurements with sound horizon based measurements improves further and the new best fit value of $H_0$ becomes  $H_0=68.3 \pm 0.5$km s$^{-1}$ Mpc$^{-1}$.

The GW+kilonovae subset combines measurements based on gravitational waves with measurements based on kilonovae. Even though this subset encompasses diverse methods, it shows good internal consistency (reduced $\chi^2 = 1.16106$) and yields a best-fit $H_0$ value of $66.3 \pm 2.0$ km s$^{-1}$ Mpc$^{-1}$, which is in close agreement with the Planck 18 result. This supports the potential of gravitational wave standard sirens as a promising independent probe of $H_0$.

The Lensing subset, after excluding the TDCOSMO I outlier, demonstrates improved consistency (reduced $\chi^2 = 0.892011$) and a best-fit $H_0$ of $69.7 \pm 1.2$ km s$^{-1}$ Mpc$^{-1}$, aligning more closely with other independent measurements.

Cosmic Chronometers and the revised Lensing subset both show excellent internal consistency (reduced $\chi^2 < 1$) and yield $H_0$ values that are in good agreement with the overall trend of one-step measurements.

The Megamasers subset, while showing some tension (reduced $\chi^2 = 1.38689$), is based on only two data points. The tension is apparently due to the outlier MCP+SH0ES datapoint\cite{Pesce:2020xfe} discussed above.

{\bf The outlier data:}
Here we discuss in some more detail the two outlier one-step $H_0$ measurements and point out potential sources of systematic errors in these measurements.
\paragraph{TDCOSMO.I measurement (2019) \cite{Millon:2019slk} $H_0= 74.2 \pm 1.6$}: The TDCOSMO I measurement (point 3 of Table II) suffers from systematic issues related to the mass-sheet degeneracy (MSD). This early analysis by \cite{Millon:2019slk}  assumed simple parametric mass models (power-law or composite) for the lens galaxies, which implicitly break the MSD. However, these assumptions can potentially bias the $H_0$ measurement or underestimate the errors if the true mass profile deviates from these simple models.

Subsequent analyses, particularly \cite{Birrer:2020tax} (TDCOSMO IV) and \cite{TDCOSMO:2023hni}, have effectively replaced this measurement by addressing the MSD issue more rigorously. These later studies have much larger uncertainties because they relax the assumptions on the mass profile and instead use stellar kinematics to constrain the MSD. Ref. \cite{Birrer:2020tax} (measurment 5 of Table \ref{tab:independent_h0_measurements}) used single-aperture stellar kinematics for the TDCOSMO sample, while \cite{TDCOSMO:2023hni} (measurment 19 of Table \ref{tab:independent_h0_measurements}) used spatially-resolved kinematics for RXJ1131$-$1231. By allowing more freedom in the mass model and constraining it with kinematic data, these analyses provide a more robust, albeit less precise, measurement of $H_0$. The increased uncertainties reflect the additional degeneracies introduced when the mass profile is not assumed to follow a simple parametric form.

\paragraph{MCP-SH0ES measurement (2020) \cite{Pesce:2020xfe} $H_0=73.9 \pm 3.0$}: This analysis attempts to estimate $H_0$ using 6 megamasers in galaxies with relatively low redshifts ($z \approx 0.002 - 0.034$), which are barely in the Hubble flow. Given these low redshifts, it is expected that peculiar velocities should lead to increased uncertainties in the $H_0$ measurement. However, the reported uncertainty of $\pm 3.0$ km s$^{-1}$ Mpc$^{-1}$ seems surprisingly small given this effect. Notably, this measurement is in tension with previous megamaser-based $H_0$ determinations, which found significantly lower values (e.g., $H_0 = 66.0 \pm 6.0$ km s$^{-1}$ Mpc$^{-1}$ from Gao et al. 2016 \cite{Gao:2015tqd} and $H_0 = 68 \pm 9$ km s$^{-1}$ Mpc$^{-1}$ from Kuo et al. 2013 \cite{Kuo2013}). The lack of any updated megamaser-based $H_0$ measurements since 2020 raises further questions about the robustness of this result. Other potential sources of systematics include the assumptions made in the disk modeling, particularly the treatment of warps and eccentricity, as well as possible biases in the selection of megamaser galaxies (6 chosen out of more than 15 known\cite{Gao2017}). The reliance of the analysis on external galaxy flow models to correct for peculiar velocities introduces additional model-dependent uncertainties. Furthermore, the small sample size makes the measurement susceptible to cosmic variance effects.

These arguments along with the results of Table \ref{tab:chi_square_analysis} support the robustness of our findings across various subsets of the distance ladder independent data. The consistency observed after removing known problematic outliers strengthens our conclusion that one-step methods tend to favor lower $H_0$ values, in better agreement with Planck 18 results, compared to distance ladder-dependent measurements.

\paragraph{The distance ladder sample: Correlations and over fitting}
Interestingly, the analysis of the full distance ladder dependent dataset yields a notably different result. With a best-fit $H_0$ of $72.7^{+0.5}_{-0.5}$ km s$^{-1}$ Mpc$^{-1}$, this dataset shows a clear preference for higher $H_0$ values compared to most of the independent subsets. However, the most striking feature is the unusually low reduced $\chi^2$ value of 0.52.

A reduced $\chi^2$ significantly below 1, as observed in the distance ladder dependent dataset, suggests that the model is "overfitting" the data. This could be due to several factors:
\begin{enumerate}
\item
Overestimated uncertainties: If the reported uncertainties in the distance ladder measurements are consistently larger than the true uncertainties, it would lead to an artificially low $\chi^2$ value.
\item
Correlation between measurements: If there are unaccounted correlations between different distance ladder measurements, it could lead to an underestimation of the true variance in the data, resulting in a lower $\chi^2$.
\item
Publication bias: There might be a tendency to publish results that are in agreement with previous measurements, leading to an artificially low scatter in the published values.
\end{enumerate}
This unusually low reduced $\chi^2$ in the distance ladder dependent dataset contrasts sharply with the more typical values seen in the independent dataset subsets. It suggests that the uncertainties in the distance ladder measurements may be overestimated, or that there are significant correlations or dependencies among these measurements that are not being accounted for in the analysis.

\paragraph{Interpretation and the Road Ahead:}
This analysis highlights the importance of critically examining individual measurements and their potential systematic effects. It also demonstrates the value of using multiple independent methods to constrain $H_0$, as this approach allows for the identification and mitigation of method-specific biases, systematics or new physics.

Given these results, the most probable interpretation is that the distance ladder measurements should be carefully reexamined. This reanalysis should focus on identifying potential systematics or local physics phenomena that may not have been adequately accounted for in previous distance ladder analyses. The consistent discrepancy between distance ladder methods and other independent techniques points to the likelihood of unresolved issues within the distance ladder methodology or assumptions.

The analysis presented here underscores the critical importance of diverse measurement techniques in cosmology. The agreement between one-step measurements and sound horizon based estimates provided by CMB observations (Planck 18) suggests that the resolution to the Hubble tension may lie in refining our understanding of local distance ladder measurements rather than in revising our models of the early universe.

Future efforts should concentrate on:
\begin{enumerate}
    \item Enhancing the accuracy and reducing systematic uncertainties in distance ladder methods.
    \item Further developing and refining one-step measurement techniques.
    \item Investigating potential local universe phenomena that could affect distance ladder measurements without impacting one-step or CMB-based methods.
    \item Conducting cross-checks and joint analyses between different measurement techniques to identify and resolve discrepancies.
\end{enumerate}

By pursuing these avenues, we can work towards a more coherent understanding of the universe's expansion rate and potentially resolve the long-standing Hubble tension.

\section{Conclusion-Discussion-Future Prospects}

This study presents a comprehensive analysis of Hubble constant ($H_0$) measurements, categorizing them into two distinct groups: distance ladder measurements and one-step measurements. The distance ladder approach, which relies on various calibrators and cosmic distance indicators, yielded an $H_0$ value of $72.8 \pm 0.5$ km/s/Mpc. In contrast, one-step measurements, which are independent of the CMB sound horizon scale and do not require a distance ladder, indicated a lower $H_0$ of $69.0 \pm 0.5$ km/s/Mpc. It's important to note that these uncertainties are likely underestimated, as the $H_0$ measurements were assumed to be independent, an assumption not fully realized in practice.

The discrepancy between these results suggests a significant shift in our understanding of the Hubble tension. Contrary to the prevailing view that the tension primarily exists between local measurements and early-time observations, this analysis indicates that the core of the tension lies between distance ladder measurements and all other methods. These other methods include local one-step measurements, early-time measurements independent of the sound horizon, and sound horizon-based measurements.

If this conjecture holds true, it implies that the origin of the Hubble tension may be rooted in misunderstood local physics affecting all distance ladder measurements similarly. This could manifest in two ways:

\begin{enumerate}
    \item A fundamental change in physics that impacts the relationships among the distance ladder rungs\cite{Ruchika:2024ymt,Perivolaropoulos:2022khd,Marra:2021fvf,Alestas:2021nmi,Perivolaropoulos:2021bds,Huang:2024erq,Liu:2024vlt,Desmond:2020wep,Desmond:2019ygn}.
    \item An unknown local systematic effect influencing all distance ladder calibrators in a similar manner\cite{Wojtak:2015gra,Wojtak:2022bct,Efstathiou:2020wxn,Carneiro:2022rmw,Mortsell:2021nzg,Rameez:2019wdt,Gialamas:2024lyw,Chen:2024gnu}.
\end{enumerate}

Thus, the present analysis offers a novel perspective on the Hubble tension, suggesting that the discrepancy may primarily lie between distance ladder measurements and other methods. This finding opens up several avenues for future research and potential extensions of our work:

\begin{enumerate}

\item \textbf{Probing Physics Changes in Distance Ladder Regimes:}
A crucial extension would involve a comprehensive search for evidence of physics changes in astrophysical observations that overlap in redshift space with the rungs of the distance ladder, particularly the first two rungs\cite{Alestas:2021nmi,Paraskevas:2023aae}. This could include:
\begin{itemize}
\item Analyzing gravitational lensing data in the relevant redshift ranges for any anomalies.
\item Investigating galaxy rotation curves and galaxy cluster dynamics for potential deviations from expected behavior.
\item Examining the behavior of standard candles and standard sirens across different distance scales\cite{Gomez-Valent:2023uof,Paraskevas:2023aae,Dainotti:2023ebr,Perivolaropoulos:2022khd,Perivolaropoulos:2021bds}.
\end{itemize}

\item \textbf{Theoretical Model Construction:}
Developing theoretical models that predict localized changes in physics, especially gravitational physics, could provide a framework for understanding potential discrepancies\cite{Perivolaropoulos:2022txg,Desmond:2020wep,Desmond:2019ygn}. These models might include:
\begin{itemize}
\item Modified gravity theories with scale-dependent effects \cite{Perivolaropoulos:2022txg,Desmond:2020wep,Desmond:2019ygn,Hogas:2023pjz}.
\item Models incorporating interactions between dark energy and standard model particles that become significant at specific scales.
\item Theories proposing variations in fundamental constants over cosmic time or distance.
\end{itemize}

\item \textbf{Reanalysis of Distance Ladder Datasets:}
A critical extension would involve reanalyzing existing distance ladder datasets with additional degrees of freedom to test if the data supports more complex models. This could include:
\begin{itemize}
\item Introducing new parameters in the Cepheid calibration that allow for a change in the Period-Luminosity (P-L) relation parameters at certain redshifts or distances\cite{Perivolaropoulos:2022khd,Perivolaropoulos:2021bds}.
\item Exploring potential environmental dependencies of standard candle luminosities\cite{Wojtak:2015gra,Wojtak:2022bct,Efstathiou:2020wxn,Carneiro:2022rmw,Mortsell:2021nzg,Rameez:2019wdt}.
\item Investigating the possibility of a smooth transition in physical laws affecting distance measurements across different scales.
\end{itemize}

\item \textbf{Expansion of $H_0$ Measurement Compilations:}
To improve statistical robustness, future work could aim to expand the compilation of $H_0$ measurements. This expansion could involve:
\begin{itemize}
\item Incorporating new measurements as they become available from ongoing and future surveys.
\item Reanalyzing older datasets with updated methodologies to potentially extract more precise $H_0$ estimates.
\item Actively seeking out and including $H_0$ measurements from diverse methodologies to ensure a comprehensive representation of measurement techniques.
\end{itemize}

\item \textbf{Detailed Correlation Analysis:}
A more nuanced treatment of correlations among the measurements considered in this study could provide valuable insights. This extension would involve:
\begin{itemize}
\item Developing a comprehensive correlation matrix for all included $H_0$ measurements.
\item Performing a principal component analysis to identify the most significant independent factors contributing to the observed tensions.
\item Investigating how accounting for these correlations affects the statistical significance of the observed discrepancies between measurement methods.
\end{itemize}

\item \textbf{Systematic Error Investigation:}
A detailed exploration of potential systematic errors specific to distance ladder measurements could help identify the source of the observed discrepancy. This could include:
\begin{itemize}
\item Reanalyzing the calibration of local distance indicators, such as Cepheids and RR Lyrae stars.
\item Investigating potential biases in supernova Ia standardization techniques.
\item Examining the impact of cosmic variance on distance ladder $H_0$ measurements.
\end{itemize}

\item \textbf{Cross-Methodology Consistency Checks:}
Developing new methods to cross-check results between different $H_0$ measurement techniques could help isolate potential issues. This might involve:
\begin{itemize}
\item Creating hybrid measurement techniques that combine aspects of distance ladder and one-step methods.
\item Designing observational tests that can discriminate between different proposed explanations for the Hubble tension.
\end{itemize}
\end{enumerate}
By pursuing these extensions, we can hope to gain a deeper understanding of the nature of the Hubble tension and potentially resolve this significant cosmological puzzle. The resolution may lie in uncovering subtle systematic effects, or it may point towards exciting new physics beyond our current standard models. In either case, these investigations promise to advance our understanding of the universe's expansion history and fundamental physics.

\section*{Acknowledgements}
I thank Adam Riess for insightful comments and stimulating discussions which motivated the addition of the $\chi^2$ analysis that identified the two outliers. This article is based upon work from COST Action CA21136 - Addressing observational tensions in cosmology with systematics and fundamental physics (CosmoVerse), supported by COST (European Cooperation in Science and Technology). This project was also supported by the Hellenic Foundation for Research and Innovation (H.F.R.I.), under the "First call for H.F.R.I. Research Projects to support Faculty members and Researchers and the procurement of high-cost research equipment Grant" (Project Number: 789).

{\bf Numerical Analysis Files:} The numerical analysis files used in this work are publicly available through GitHub. These files can be accessed  \href{https://github.com/leandros11/distladder/}{through this link}. We encourage interested readers to explore this repository for further details on our computational methods and to facilitate reproducibility of our results.

\bibliography{references} 

\end{document}